\documentclass{aa}

\usepackage{graphicx}
\usepackage{natbib}
\usepackage{amsmath}
\usepackage{xcolor}
\usepackage{txfonts}
\usepackage[colorlinks=true]{hyperref}
\usepackage{ulem}
\hypersetup{linkcolor=blue,citecolor=blue,filecolor=blue,urlcolor=blue}
\graphicspath{{./}{figures/}}

\def\mbh{$M_{\bullet}$}
\def\eddr{$\alpha_{\rm Edd}$}
\def\civ{{C\sc{iv}} $\lambda$1549\/}
\def\mgii{{Mg\sc{ii}} $\lambda$2800\/}

\def\mgiia{{Mg\sc{ii}}~}
\def\feii{{Fe\sc{ii}}~}

\begin{document} 

\title{Dust-driven wind as a model of broad absorption line quasars}

\titlerunning{Dusty wind model of BAL QSOs}

   \author{M. H. Naddaf
          \inst{1,2}\fnmsep\thanks{naddaf@cft.edu.pl}
          \and
          M. L. Martinez-Aldama
          \inst{3,4}
          \and
          P. Marziani
          \inst{5}
          \and
          S. Panda
          \inst{6}$^{\thanks{CNPq Fellow}}$
          \and
          M. Sniegowska
          \inst{2,1}
          \and
          B. Czerny
          \inst{1}    
          }

  \institute{Center for Theoretical Physics, Polish Academy of                Sciences, Lotnikow 32/46, 02-668 Warsaw, Poland
        \and
             Nicolaus Copernicus Astronomical Center, Polish Academy of Sciences, Bartycka 18, 00-716 Warsaw, Poland
        \and     
             Instituto de Física y Astronomía, Facultad de Ciencias, Universidad de Valparaíso, Gran Bretaña 1111, Valparaíso, Chile
        \and  
             Departamento de Astronomía, Universidad de Chile, Casilla 36D, Santiago, Chile
        \and
             INAF-Astronomical Observatory of Padova, Vicolo dell'Osservatorio, 5, 35122 Padova PD, Italy
        \and
             Laborat\'orio Nacional de Astrof\'isica - Rua dos Estados Unidos 154, Bairro das Na\c c\~oes. CEP 37504-364, Itajub\'a, MG, Brazil
             }


 
  \abstract
   {We test the scenario according to which the broad absorption line (BAL) phenomenon in quasars (QSOs) is not a temporary stage of their life. In this scenario, the BAL effect acts only if the line of sight is within a spatially limited and collimated massive outflow cone covering only a fraction of the sky from the point of view of the nucleus. 
    }
   {The aim is to understand the theoretical mechanism behind the massive outflow in BAL QSOs, which is important for modelling the impact of quasars on the star formation rate in the host galaxy, and,  subsequently, on the galaxy evolution.
   }
   {We applied the specific theoretical model of dust-driven wind that was developed to explain broad emission lines. The model has considerable predictive power. 
   The 2.5D version of the model called failed radiatively accelerated dusty outflow (FRADO) includes the formation of fast funnel-shaped outflow from the disk for a certain range of black hole masses, Eddington ratios, and metallicities. We now interpret BAL QSO as sources that are viewed along the outflowing stream. We calculated the probabilities of seeing the BAL phenomenon as functions of these global parameters, and we compared these probabilities to those seen in the observational data. We included considerations of the presence or absence of obscuring torus.
   }
   {Comparing our theoretical results with observational data for a sample of QSOs consisting of two sub-populations of BAL and non-BAL QSOs, we found that in the model and in the data, the BAL phenomenon mostly occurs for sources with black hole masses higher than $10^8 M_{\odot}$. The effect increases with accretion rate, and high metallicities are also more likely in QSOs showing BAL features if a torus is taken into account.}
   {The consistency of the model with the data supports the interpretation of the BAL phenomenon as the result of the orientation of the source. It also supports the underlying theoretical model, although more consistency checks should be made in the future.
   }


   \maketitle


\section{Introduction}
\label{sec:intro}

Broad absorption line quasars (BAL QSOs) were discovered about half a century ago \citep{Lynds_1967}. The peculiar blue-shifted absorption features in their spectra show the clear signature of massive energetic outflows from the underlying source \citep[see e.g.][]{weymannetal91} with typical outflow velocities of about several up to tens of 1000 km/s, but reaching 0.2 c in the most extreme cases \citep[e.g.][]{Risaliti2005, Hidalgo2022}.
The massive outflows from BAL sources and their relatively wide opening angle can play an important role in the feedback process in active galaxies and its consequences on the evolution of the central black hole and galactic bulge \citep[see e.g.][]{DiMatteo2005, Elvis2006, Hopkins2009, Moe2009, hamann2019}.

Depending on the studied sample, the BAL phenomenon is present in 
about 5\% to  even 40\% of the total quasar population \citep[see e.g.][]{weymannetal91, Tolea_etal_2002, Knigge_etal_2008, Allen_etal_2011, Shen_etal_2019}. A very recent study reported an even higher BAL fraction of 47\% \citep{bischetti2022}. Observationally, however, we cannot distinguish between the two basic scenarios describing the BAL quasar phenomenon:

(i) The BAL phenomenon is a short stage in the quasar life cycle, and the outflow is spherically symmetric \citep[e.g.][]{surdejswing81,drewboksenberg84,voit1993,krolikvoit98}.

(ii) The BALs occur in all QSOs showing signatures of outflow, but BAL are only detected if our line of sight is within the outflow cone that is collimated and spatially limited. This cone covers only a fraction of the sky from the point of view of the nucleus \citep{young1999, elvis2000, Elvis2012radiation}.

Both scenarios predict that only a fraction of QSOs shows BAL features, and they cannot be distinguished on the basis of statistical grounds alone, although scenario (i) is no longer favoured \citep[][and the references therein]{turnshek88}.

In this study, we follow the second scenario, but we use a theoretical framework that allows us to predict the outflow cone as a function of quasar parameters. Compared with available statistics, this can tell us whether our physically motivated model is a viable explanation of the BAL phenomenon.

Scenario (ii) is usually formulated as an empirical picture. \citet{elvis2000, Elvis2012radiation} proposed a likely line-driven funnel-shaped outflow with an inclination of about 60 degrees with respect to the symmetry axis and an opening angle of 6 up to 12 degrees. The corresponding covering factor of the funnel, which is between 0.1 to 0.2, is then assumed as the BAL prevalence or the probability of observing the BAL effect, equivalently referring to the 10 to 20\% population of BAL QSOs to all quasars. The geometry is purely empirical. No physical parameters are included to constrain the geometry of the outflow, but the scenario accounts for all emission and absorption features.

This picture, however, does not account for all BAL properties. The outflow in the picture of \cite{elvis2000} is assumed to be powered by a line-driving mechanism \citep[see e.g.][]{nomura2013}, while studies of BAL objects imply the presence of dust with a reddening curve characteristic of predominately large grains \citep[see e.g.][]{dunn2010, bautista2010}, hence making dust-based models of the outflow crucial. Moreover, the range of distances for the location of the outflow is reported to be much greater \citep[see e.g.][]{korista2008, dunn2010, borguet2013, chamberlain2015} than acceleration region of winds that is frequently obtained based on the line-driving mechanism \citep{murray1995, proga2004, waters2021}. This therefore favours the dust-driving scenario, which works at radii that are larger by a factor of $\sim 10$.

In this paper, we therefore test the theoretical model of dusty winds, which was developed to explain the broad emission lines \citep{czerny2011, naddafczerny2022}, against the observational data for BAL QSOs.
The failed radiatively accelerated dusty outflow model, known as FRADO, works at the basis of a dust-driving mechanism in which disk radiation pressure acting on dust lifts up the clumpy dusty material that is initially in Keplerian orbits around the central black hole, from the disk surface. After it is lifted, the material may either follow a closed loop and fall back onto the disk surface, or it may gain velocities that are high enough to escape the gravitational potential of the central black hole towards the torus. The material depending on the trajectory may either lose the dust content because it can be irradiated by intense central disk radiation if it reaches high altitudes, or it remains dusty for the full time of flight.

The dust-driving mechanism therefore gives rise to the formation of a funnel-shaped stream of material \citep{naddaf2021}, consistent with the empirical model of \cite{elvis2000}. However,
the model gives specific predictions for the outflow geometry, depending on the black hole mass, Eddington ratio, and the metallicity of the disk material.  As in the previous approach, we expect the BAL phenomenon when the viewing angle towards the nucleus is within the outflow cone. We aim to determine whether the theoretically predicted trends with quasar parameters are consistent with observations of BAL QSOs.

The paper is therefore structured as follows. We introduce the sample of BAL QSOs we selected for the purpose of this study in Section \ref{sec:data}. The theoretical model of dusty winds describing the geometry of a massive outflow along with the grid of simulations is introduced in Section \ref{sec:theory}. The results are presented in Section \ref{sec:results}, followed by a discussion in Section \ref{sec:discussion}.

In the following, we use the definitions of the basic parameters for a single source,
\begin{equation}\nonumber
    M_{\bullet} =  M_{\rm BH} / M_{\odot}~,~~~
    \dot{m} = \dot{M} / \dot{M}_{\rm Edd}~,~~~
    \alpha_{\rm Edd} = L / L_{\rm Edd},
\end{equation}
where $M_{\bullet}$ is the dimensionless mass of the black hole in solar units, $L_{\rm Edd}$ is the Eddington luminosity, $\dot{M}_{\rm Edd} = L_{\rm Edd} c^{-2} \eta^{-1}$ is the Eddington accretion rate (where the radiative efficiency of $\eta = 0.1$ is adopted), and $\dot{m}$ and $\alpha_{\rm Edd}$ are the dimensionless accretion rate and luminosity of the source in Eddington units, respectively.

\section{Selected BAL QSO observational sample}
\label{sec:data}

\begin{figure}[b!]
    \centering
    \includegraphics[width=0.82\columnwidth]{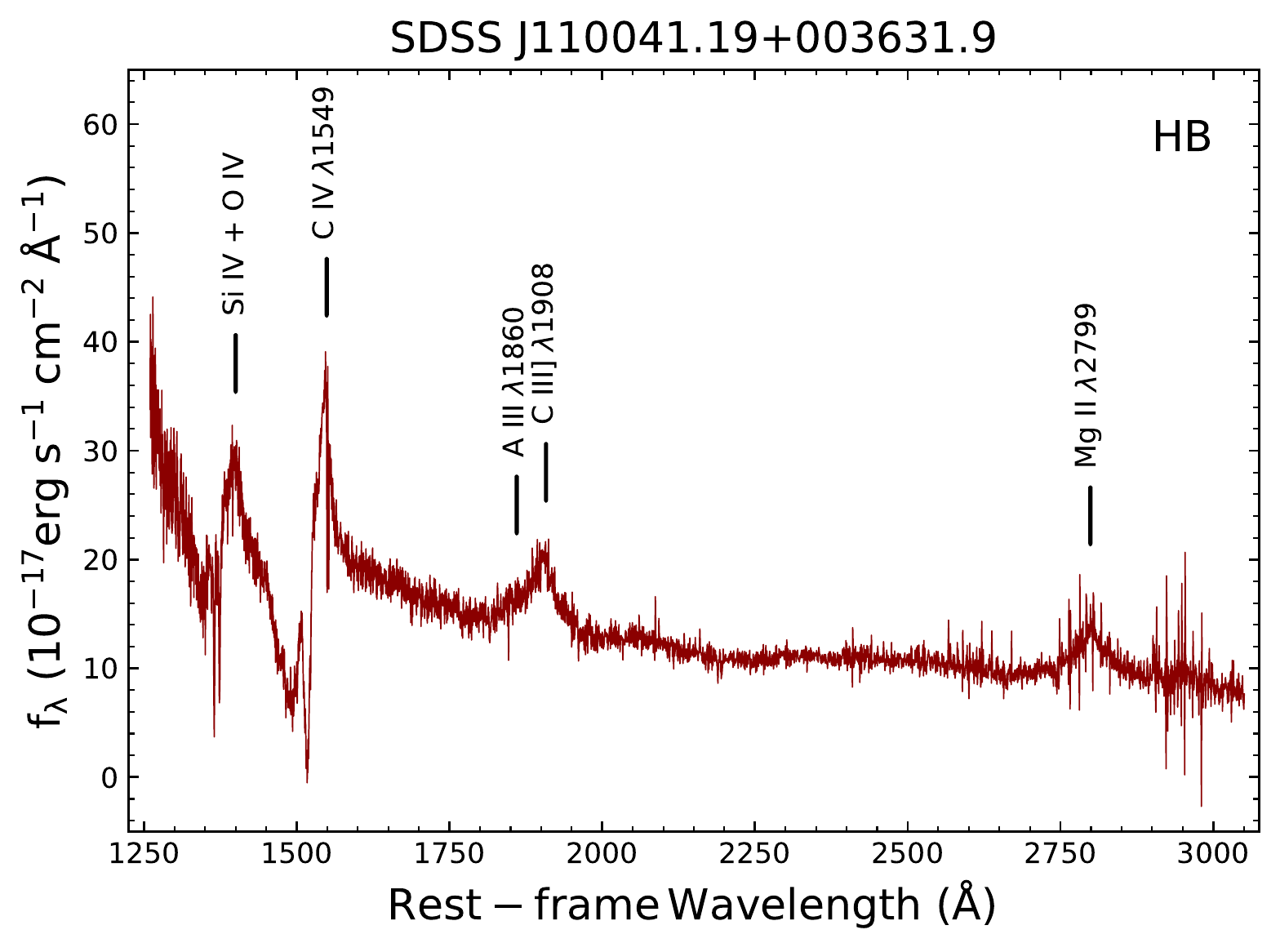}\\
    \vspace{6mm}
    \includegraphics[width=0.82\columnwidth]{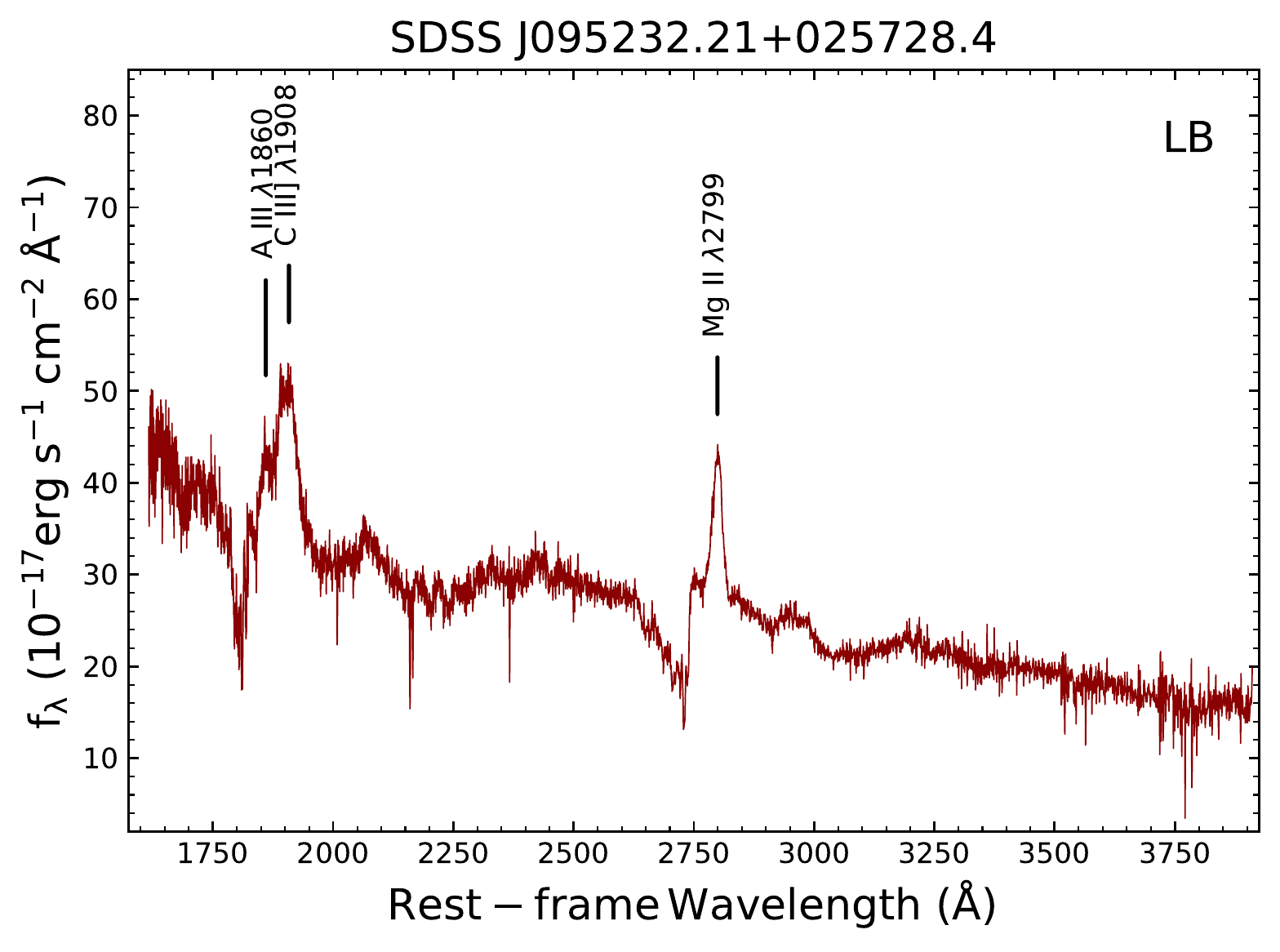}\\
    \vspace{3mm}
    \caption{Representative SDSS spectrum for an HB (upper panel) and an LB (lower panel) from our working sample. The spectra have been scaled to rest-frame wavelengths. The prominent emission lines are marked on the spectra.}
    \label{fig:sdss-spectra}
\end{figure}

\begin{figure*}[t!]
  \centering  
    \includegraphics[width=\textwidth]{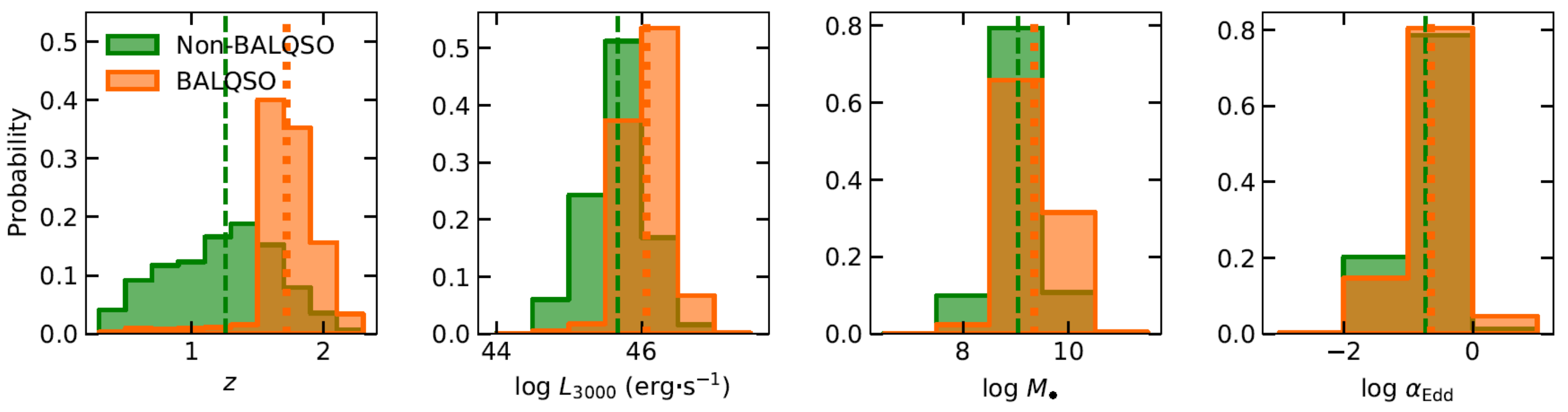}
    \caption{Redshift ($z$), luminosity at 3000$\AA$ ($L_{3000}$), black hole mass (\mbh), and Eddington ratio (\eddr) distributions for the non-BAL QSO (red,) and BAL QSO (green) samples selected from the \citet{shen2011} catalogue. The vertical dashed and dotted lines indicate the median values for non-BAL QSO and BAL QSO, respectively.}
    \label{fig:histograms}
\end{figure*}

For the current study, we used the quasar catalogue of the Sloan Digital Sky Survey DR7 \citep{shen2011}. This catalogue provides an identification of the \civ\ high and \mgii\ low BAL QSO, respectively. 
We performed a visual inspection to ensure the correct identification of the BAL systems.  The objects were classified according to their spectral appearance in a high BAL QSO (HB) showing only the absorption transitions from highly ionised atoms, a low BAL QSO (LB) showing both the high-ionisation transitions and absorption lines from lower-ionisation ions in their rest-UV spectra, a mini-BAL QSO (MB), and an FeLow BAL QSO (FLB) that shows \feii absorption lines. Representative spectra for HB and LB are shown in Fig.~\ref{fig:sdss-spectra}.

The MBs were excluded because they are characterized by systematically narrower troughs that are often associated with blended narrow-line absorptions, and their Balnicity index is 0 \citep{weymannetal91}. The MB phenomenology suggests that they belong to a different physical scenario from that of LB and HB \citep{vestergaard03,sulenticetal06a}. An FLB is thought to represent an extreme aspect of  the BAL phenomenon \citep{choietal22,choietal22a}; however, due to the complex appearance of their spectra, which are strongly affected by deep and broad \feii absorptions throughout the UV range, the relevant physical parameters and even redshifts are poorly known. The FLBs we identified are a small subsample and were excluded from the analysis of the present work because dedicated work is needed to properly ascertain their properties. This goes beyond the scope of the present analysis.

The sample includes sources with a median S$/$N$>10$ per pixel in the rest-frame 2700-2900${\rm \AA}$ region \mgiia\ \citep[see column 102 of table 1 in][]{shen2011} and black hole masses estimated from \mgii. The latter condition is necessary because in the UV range, this emission line is a better virial estimator than \civ\ \citep[e.g.][]{mejia-restrepo16}.  The full sample includes 42,349 objects where 40,991 and 1,358 are non-BAL QSO and BAL QSO (HB and LB), respectively. The \mgiia\ black hole mass and the Eddington ratio were taken from the SDSS catalogue  \citep[see columns 135 and 141 of table 1 in][]{shen2011}.
 
To summarise, the sample spans  a redshift, the luminosity at 3000\AA, the black hole mass, and Eddington ratio ranges of $0.3501<z<2.2487$, $45.04<{\rm log }L_{3000}<48.19$, $7.28<{\rm log } M_{\bullet~\rm MgII}<11.31$, and $-2.88<{\rm log } \alpha_{\rm Edd}<0.69$, respectively. Fig.~\ref{fig:histograms} shows the distributions of $z$, $\log L_{3000}$, $\log M_\bullet$, and $\log \alpha_\mathrm{Edd}$\ for the BAL QSO and non-BAL QSO samples. The Eddington ratio for BAL and non-BAL samples seem similar, but the masses in BAL quasars are higher on average, and (consistently) luminosities are also higher on average. BAL quasars seem to be located systematically at higher redshifts, which might also reflect the fact that statistically, we detect higher-mass quasars at higher redshifts \citep[e.g.][]{vestergaard2006}. {According to the $p$-values ($\ll 0.01$) of a Kolmogorov-Smirnov (KS) two-sided test, all the distributions are drawn from different parent populations. 

\section{Theoretical model}
\label{sec:theory}

The 2.5D FRADO model \citep{naddaf2021}, which is a relatively simple non-hydrodynamical approach to the dynamics of the lowly ionised part of a BLR, leads to the formation of a fast outflow from the disk for high black hole masses, Eddington ratios, and metallicities, although it is based on a radiatively dust-driving mechanism and addresses the farther part of the BLR that is located at about a few thousand gravitational radii ($r_g$), where the disk is cold enough for the dust to form. The basic FRADO model \citep{czerny2011,czerny2015,czerny2017} describing the dynamics of BLR material in one dimension, that is, vertical with respect to the disk plane, could not address the outflow feature, but the 2.5D enhanced version of the model \citep{naddaf2021} predicted the emergence of this feature for a certain range of the global source parameters.

The predictions of the 2.5D version of dust-driven outflow from the accretion disk were recently tested against observational data with the radius-luminosity relation \citep{naddaf2020} and the shape of observed broad emission lines in quasars \citep{naddafczerny2022}, which altogether make the model an interesting case for further studies. Therefore, we intend to investigate the geometry of the outflow originating from the accretion disk in the lowly ionised part of the BLR based on the 2.5D FRADO model. 

\subsection{Dust-driving mechanism in 2.5D FRADO}
\label{sec:mechanism}

The basic FRADO model analytically only described the failed dusty wind motion and did this only in a 1D approximation (vertical motion with respect to the disk plane), folding the radiation flux of the accretion disk with the wavelength-averaged dust opacity \citep{czerny2011, czerny2015, czerny2016, czerny2017}. Despite many interesting results in explaining the geometry and physics of the broad line region, it was not able to catch the dynamical picture of a likely expected outflowing wind. 

In 2.5D FRADO, the clumpy dusty material, initially in Keplerian orbits around the central black hole, is launched due to radiation emitted by the whole extended disk acting on dust at the disk surface. Radiation is geometrically moderated by the shielding, which depends on the cloud location \citep[see][for details]{naddaf2021}. Known as shielding effect, it is a requirement, protecting the early lifted clumps from too early sublimation by the intense central disk radiation that is necessary to launch an efficient outflow \citep[see e.g.][]{murray1995, risaliti2010, naddaf2021}.
The radiation pressure force was calculated allowing for wavelength-dependent dust opacities for a range of dust grains as in \citet{szczerba2013}, and the assumed dust distribution came from \citet{mathis1977}. The radiative force was thus calculated by convolving the locally available radiation flux, including its spectral shape, with the wavelength-dependent cross section for dust scattering and absorption (for more details, see \citealt{naddaf2021}). 

In 2.5D FRADO, the full pattern of the trajectories of clumpy dusty-gaseous material launched at different radii of the disk is complex and depends on the main global parameters of the source, such as mass, Eddington rate, and metallicity. They determine the launching radius, and if the available radiative force is not strong enough, the lifted material later returns to the disk surface, hence forming a failed wind. If the available radiative force is high enough, the material will escape toward the torus in the form of a fast outflowing stream of material. The dust sublimation temperature was fixed at 1500 K for all dust grains \citep{baskin2018, Gravity2020}, which consequently set a geometrical location above the accretion disk, the so-called sublimation location, as depicted with the solid curved dark red line in Figure \ref{fig:schematic}, where the material crossing it loses the dust content and otherwise remains dusty. The geometry of the lowly ionised part of the BLR, including the outflowing wind, in the 2.5D FRADO model, is thus determined by the kinematics of the clouds, which in turn is set by global source parameters. For more details, we refer to our previous papers \citep{naddaf2021, naddafczerny2022}.

We also note that general relativistic effects are negligible at the large radii that are appropriate for a lowly ionised BLR, and therefore, they are currently neglected in the model.

\subsection{Outflow geometry and probability of the BAL effect}

When the source parameters allow launching the escaping wind, a geometrically funnel-shaped structure appears, as sketched in Figure \ref{fig:schematic}. The radial range from which the launched material contributes to funnel formation is called the escaping zone.

In order to determine the probability of the BAL effect in a single source, we calculated the 3D trajectories of the material, and we determined whether we identified the escaping zone for the adopted global parameters. If this was the case, we determined the minimum and maximum of the viewing angle for the escaping trajectories, and we calculated the corresponding solid angle with respect to the nucleus. The ratio of this solid angle to half of the sphere measures the probability of observing the BAL effect in the source. Likewise, when a torus is taken into account, the partial solid angle of the outflow is not obscured by the torus. We divided it by the part of the semi-sphere that does not subtend the torus. The probability of observing a BAL is then given as

\begin{equation}\label{eq:probability}
    \Omega_{\theta_{\rm t}} = \dfrac
    {\cos{ [\, \min (\theta_2 \mid \theta_{\rm t})\,]} ~-~
     \cos{ [\, \min (\theta_1 \mid \theta_{\rm t})\,]}}
    {1 - \cos \theta_{\rm t}}
,\end{equation}
where $\theta_1$ and $\theta_2$ are the maximum and minimum viewing angles of the outflowing funnel, and $\theta_{\rm t}$ is the opening angle of the torus. All angles are measured with respect to the azimuthal axis.

The solid angle of the funnel of outflow strongly depends on all the source global parameters, that is, the black hole mass, Eddington ratio, and metallicity. It is merely determined on the basis of the dust launching, and although eventual further acceleration of the launched material by line-driving will clearly change the velocities, it is not expected to affect the direction of motion strongly. Therefore, our model is thought to represent the geometrical aspect well.

\begin{figure}[b!]
        \centering
        \includegraphics[width=0.95\columnwidth]{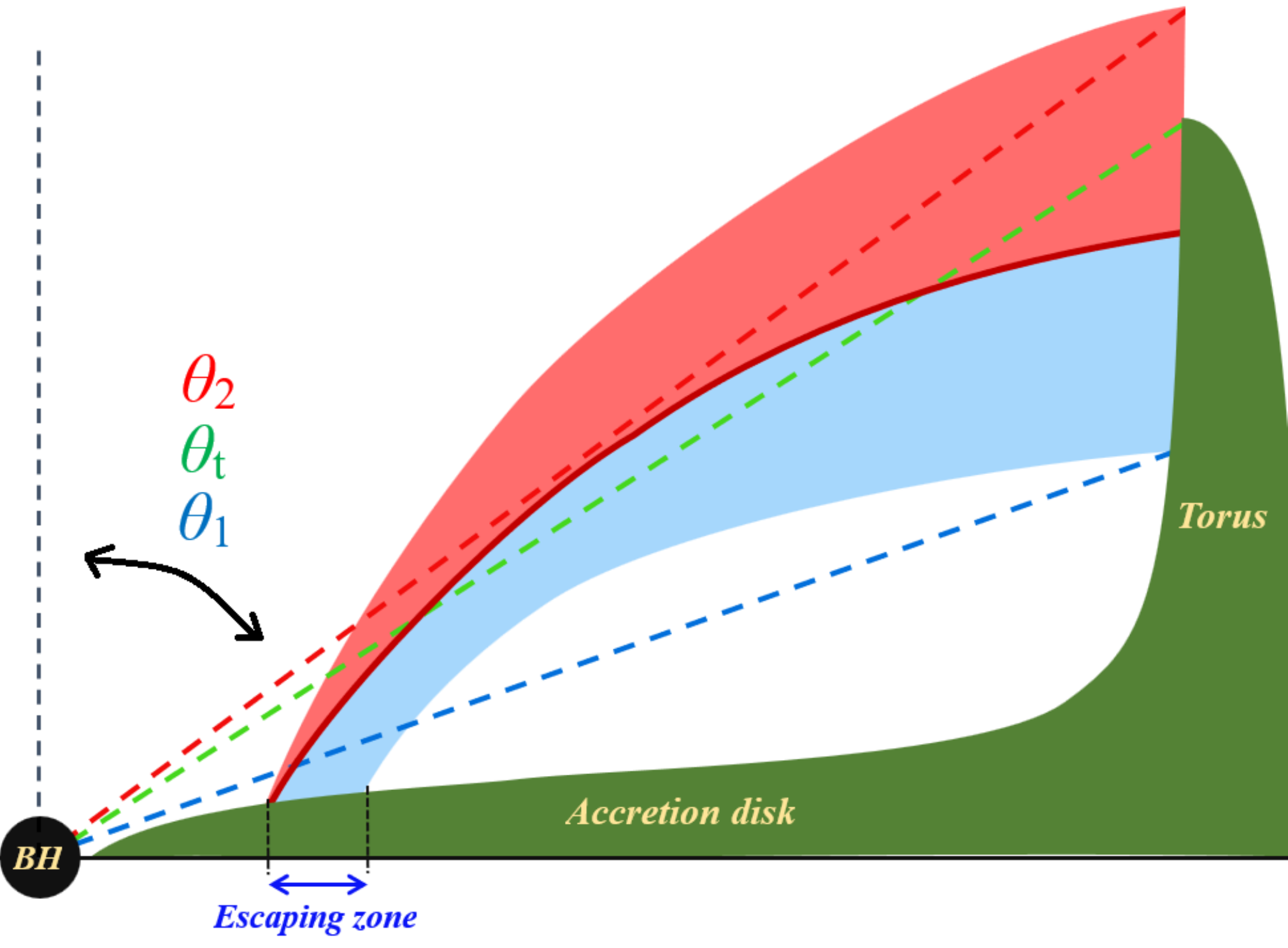}
        \vspace{2mm}
        \caption{Schematic 2D-projected illustration of a massive outflow from the accretion disk due to the radiatively dust-driving mechanism. The radial extension of the LIL BLR is the entire radial range from which the material can be lifted from the disk surface, and the escaping zone represents the radial range from which all lifted material escapes to the torus or infinity. The dashed lines indicate the reference angles with respect to the azimuthal symmetry axis, i.e. the opening angle of the torus $(\theta_{\rm t})$, the maximum angle $(\theta_1),$ and the minimum  angle $(\theta_2)$ of the outflow funnel (cone), which are shown in green, blue, and red, respectively. The solid curved line in dark red represents the sublimation location in which the material that crosses loses the dust content. The regions in the funnel-shaped wind structure shaded in blue and red correspond to dusty and dustless escaping material, respectively.}
        \label{fig:schematic}
\end{figure}

\subsection{Grid of models}
\label{sec:setup}

We computed the probability of the BAL phenomenon for a grid of three main parameters. The black hole mass, $M_{\bullet}$, was assumed to be between 6 and 10, on a logarithmic scale with a step size of one. For the dimensionless accretion rate, $\dot{m}$, we used four values: -2, -1, 0, and 1, also in the logarithmic scale. Finally, we considered values of the metallicity on the linear scale: 1, 2.5, 5, and 10, in solar units.

The range of masses for the central black hole and the range of accretion rates were adopted in order to cover not only the domain of BAL QSOs available in the sample, but also the reasonable entire range of the observed QSOs and even lower-mass AGN. We did not consider lower accretion rates because then the flow towards the black hole might not always continue in the form of a geometrically thin optically thick disk \citep[for AGN properties, see][]{krolik_book_1999}. We considered only solar and super-solar metallicities because most studies imply this metallicity range \citep[e.g.][]{juarez2009,sniegowska2021,garnica2022}.

For each model, we calculated the trajectories for a very dense set of initial launching radii in order to have a high-resolution picture of the outflow geometry, if it forms, and we then determined the parameters of the outflow cone.

The computation results for the torus are subject to further processes with considered torus opening angles of 45, 60, 75, and 90 (no torus) degrees. These values were selected in such a way as to include the considered extreme modes of obscuration by a torus and also two cases of mild obscuration.

Notably, the dust sublimation temperature was kept at 1500 K in all cases, but the overall picture of the outflow geometry including the opening angle and orientation is not expected to be affected significantly, although the location of the escaping zone is then shifted accordingly.

\section{Results}\label{sec:results}

\begin{figure}[b!]
        \centering
        \includegraphics[width=\columnwidth]{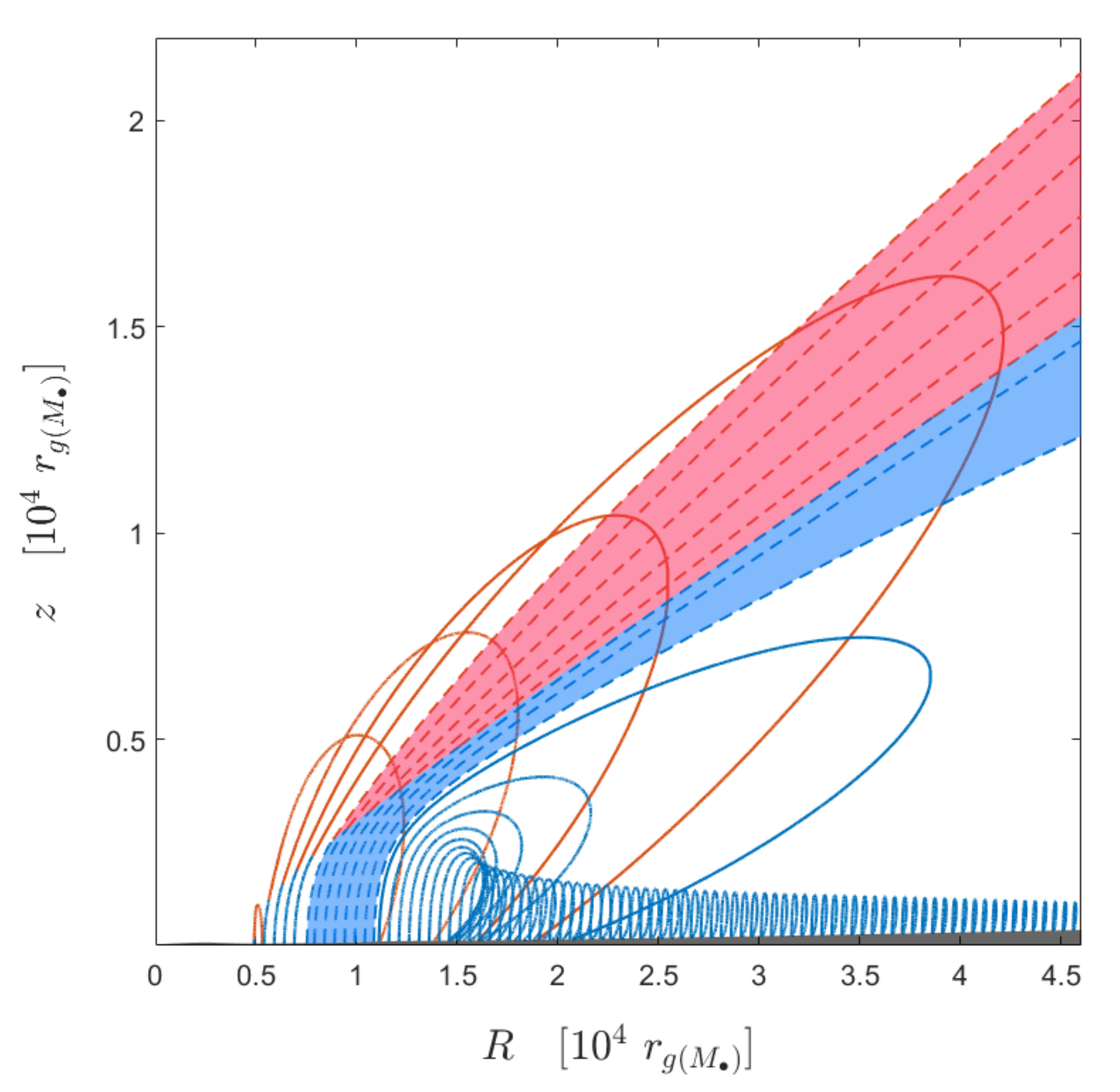}
        \caption{Example of the trajectories of clouds above the accretion disk for the case of $\log{\dot{m}}=0$, $\log{M_{\bullet}}=8$, and $Z=5 Z_{\odot}$. The motion is displayed in a 2D plane of $(R, z)$ in cylindrical coordinates due to the azimuthal symmetry, where $R$ is the radial distance to the black hole in the equatorial plane, and $z$ is the vertical distance from the equatorial plane. The overall motion consists of a failed wind (solid curved lines) and an outflow (dashed lines) lines). Dusty and dustless situations are colour-coded in blue and red, respectively. Only a fraction of trajectories is shown for clarity.}
        \label{fig:motion example}
\end{figure}

\subsection{Model predictions}

We first calculated the grid of theoretical models as outlined in Section~\ref{sec:setup}. For each model, a dense grid of trajectories was calculated, which allowed for the precise determination of the geometry. An example is shown in the 2D projected perspective in Figure~\ref{fig:motion example}. Their many trajectories of dusty (blue) and dustless (red) material form a failed wind, that is, material returns to the disk, but lands at a larger radius than the launch radius. However, for the present paper, the escaping trajectories are of key interest. They are marked with light blue and red belts, which form a thick funnel-shaped outflow. The light blue part remains dusty, while in the red part, dust was evaporated and the clouds there performed just a ballistic motion, but were already accelerated beyond the escape velocity. 
The line-driving mechanism may increase the opening angle of the outflow and boost the terminal velocities, but it is not yet incorporated into the model, and therefore, only the dust-driving mechanism accounts for this.
For the parameters adopted in Figure~\ref{fig:motion example}, the escaping stream forms.  We therefore list its parameters in Table~\ref{tab:compdata}.

\begin{table}[b!]
    \caption{Geometrical properties of the outflow}
    \centering
\begin{tabular}{lllccc}
\hline
$\log{\dot{m}}$ & $\log{M_{\bullet}}$ & $Z/Z_{\odot}$ &  $\theta_{1}$ & $\theta_{2}$ & $\Omega_{\rm full}$ \\
\hline \\
        $1$ & $8$ & $1$ & $69.68$ & $79.13$ & $0.1587$ \\
        $ $ & $ $ & $2.5$ & $55.96$ & $80.12$ & $0.3882$\\
        $ $ & $ $ & $5$ & $46.70$ & $79.94$ & $0.5111$ \\
        $ $ & $ $ & $10$ & $25.34$ & $81.31$ & $0.7527$ \\
\\
        $ $ & $9$ & $1$ & $54.61$ & $78.93$ & $0.3871$ \\
        $ $ & $ $ & $2.5$ & $35.76$ & $79.85$ & $0.6352$\\
        $ $ & $ $ & $5$ & $23.95$ & $83.12$ & $0.7941$ \\
        $ $ & $ $ & $10$ & $19.73$ & $85.99$ & $0.8714$\\
\\
        $ $ & $10$ & $1$ & $50.96$ & $78.97$ & $0.4385$ \\
        $ $ & $ $ & $2.5$ & $21.66$ & $81.83$ & $0.7873$\\
        $ $ & $ $ & $5$ & $18.35$ &84.12$ $ & $0.8467$ \\
        $ $ & $ $ & $10$ & $15.48$ & $85.78$ & $0.8901$\\
\hline \\
        $0$ & $6$ & $5$ & $73.67$ & $80.02$ & $0.1079$ \\
        $ $ & $ $ & $10$ & $71.39$ & $79.67$ & $0.1398$ \\
\\
        $ $ & $7$ & $2.5$ & $72.96$ & $80.02$ & $0.1197$ \\
        $ $ & $ $ & $5$ & $71.28$ & $79.60$ & $0.1404$ \\
        $ $ & $ $ & $10$ & $67.40$ & $79.15$ & $0.1961$ \\
\\
        $ $ & $8$ & $1$ & $75.48$ & $79.93$ & $0.0759$ \\
        $ $ & $ $ & $2.5$ & $71.16$ & $79.55$ & $0.1415$ \\
        $ $ & $ $ & $5$ & $66.66$ & $79.06$ & $0.2064$ \\
        $ $ & $ $ & $10$ & $54.41$ & $78.55$ & $0.3835$ \\
\\
        $ $ & $9$ & $1$ & $71.40$ & $79.65$ & $0.1393$ \\
        $ $ & $ $ & $2.5$ & $68.51$ & $78.83$ & $0.1726$ \\
        $ $ & $ $ & $5$ & $57.08$ & $77.30$ & $0.3236$ \\
        $ $ & $ $ & $10$ & $46.89$ & $78.11$ & $0.4774$ \\
\\
        $ $ & $10$ & $1$ & $67.75$ & $78.72$ & $0.1830$ \\
        $ $ & $ $ & $2.5$ & $56.10$ & $78.17$ & $0.3527$ \\
        $ $ & $ $ & $5$ & $45.89$ & $76.17$ & $0.4570$ \\
        $ $ & $ $ & $10$ & $37.63$ & $76.80$ & $0.5636$ \\
\hline \\
        $-1$ & $7$ & $10$ & $75.16$ & $79.87$ & $0.0802$ \\
\\
        $ $ & $8$ & $5$ & $74.08$ & $80.32$ & $0.1061$ \\
        $ $ & $ $ & $10$ & $71.50$ & $79.28$ & $0.1313$ \\
\\
        $ $ & $9$ & $2.5$ & $75.29$ & $80.21$ & $0.0839$ \\
        $ $ & $ $ & $5$ & $71.33$ & $78.08$ & $0.1136$ \\
        $ $ & $ $ & $10$ & $66.32$ & $78.99$ & $0.2106$ \\
\\
        $ $ & $10$ & $1$ & $77.61$ & $78.69$ & $0.0184$ \\
        $ $ & $ $ & $2.5$ & $71.84$ & $78.58$ & $0.1137$ \\
        $ $ & $ $ & $5$ & $66.98$ & $78.65$ & $0.1943$ \\
        $ $ & $ $ & $10$ & $58.88$ & $78.17$ & $0.3118$ \\
\hline \\
        $-2$ & $9$ & $10$ & $77.06$ & $80.39$ & $0.0570$ \\
\\
        $ $ & $10$ &  $5$ & $76.82$ & $78.09$ & $0.0216$ \\
        $ $ & $ $ & $10$ & $71.57$ & $79.26$ & $0.1298$ \\
\hline
\end{tabular}
    \label{tab:compdata}
\end{table}

The first three columns of Table~\ref{tab:compdata} from left to right show the accretion rate of the disk, the mass of the central black hole, and the metallicity of the disk, respectively. The fourth and fifth columns represent the minimum and maximum viewing angles of the outflow measured with respect to the symmetry axis, respectively. The last column shows the probability of observing the BAL effect based on Equation~\ref{eq:probability} without considering the obscuring torus, that is, $\theta_{\rm t}$ is set at 90 degrees, hence named $\Omega_{\rm full}$.

\begin{figure*}[t!]
        \centering
        \includegraphics[width=0.9\textwidth]{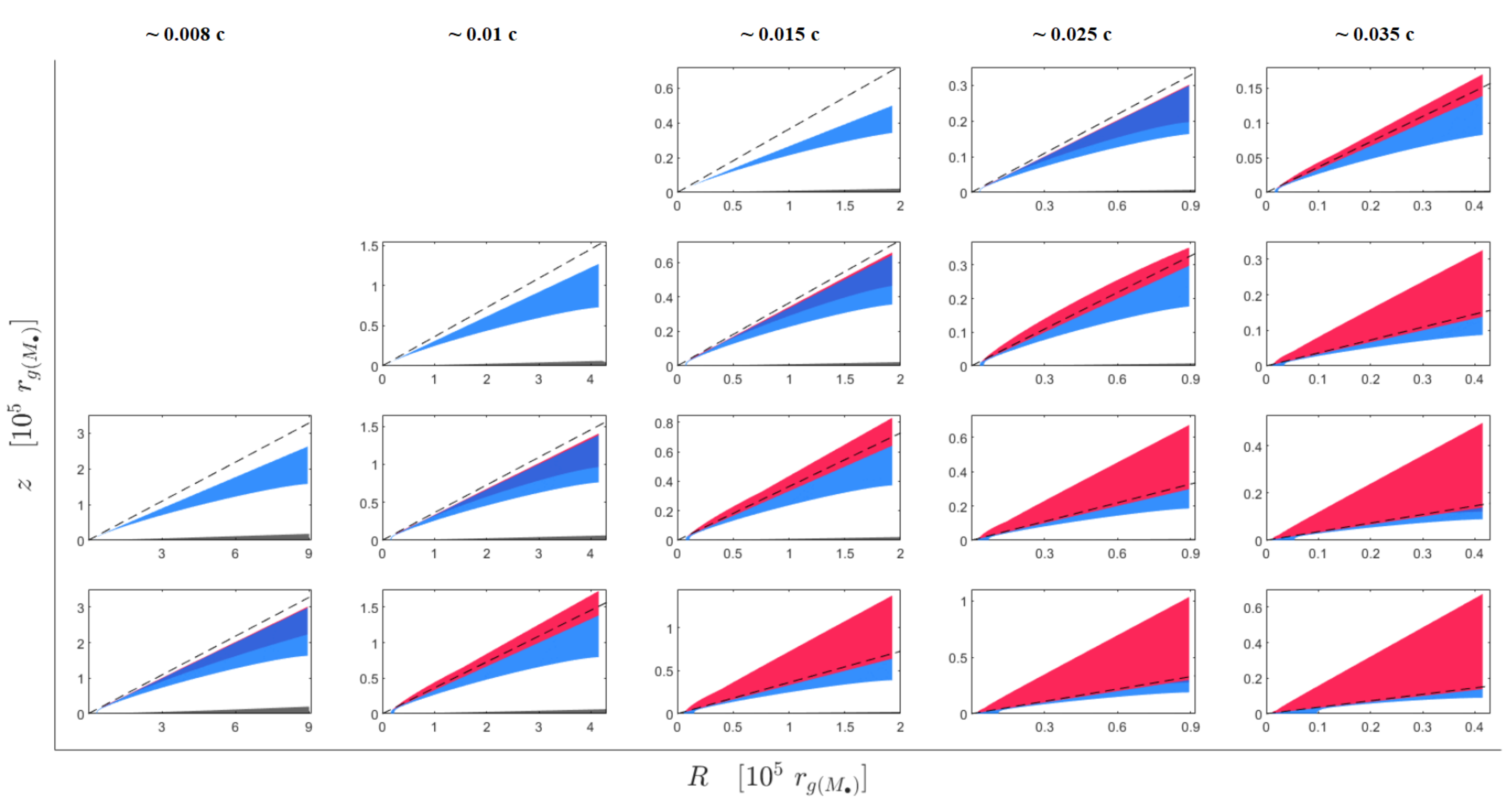}
        \caption{Results of simulations showing the massive outflow triggered by the disk radiation pressure for $\dot{m}=1$. $M_{\bullet}$ is set at 6, 7, 8, 9, and 10 (in log units) from left to right. The adopted metallicities are 1, 2.5, 5, and 10 $Z_{\odot}$, from top to bottom. The mean value of the escape velocity in the escaping zone is indicated as a reference at the top of each column. The failed wind is not shown for better contrast. The accretion disk is shown as the shaded grey area. The dusty and dustless components of the outflow funnel are shaded in blue and red, respectively. The dashed black line represents the arbitrary obscuration angle of 70 degrees by the torus as a reference.}
        \label{fig:one_Edd_outflow}
\end{figure*}

As shown in Table~\ref{tab:compdata}, the maximum viewing angles of the outflow in all cases tend to be about 81 degrees, with a spread of about 10 degrees, that is, $\theta_{2} = 81 \pm 5$. However, for the minimum viewing angles of the outflow, $\theta_{1}$, the broad set of values ranges from $\sim 15$ degrees up to $\sim 78$ degrees, depending on the main global parameters of the source. The opening angle of the outflow can therefore be as low as a few degrees and up to $\sim 70$ degrees. With the increase in black hole mass, accretion rate, and metallicity, the values of $\theta_{1}$ clearly decrease, yielding a wider opening angle for the outflow.

\begin{figure*}[h!]
        \centering
        \includegraphics[width=0.9\textwidth]{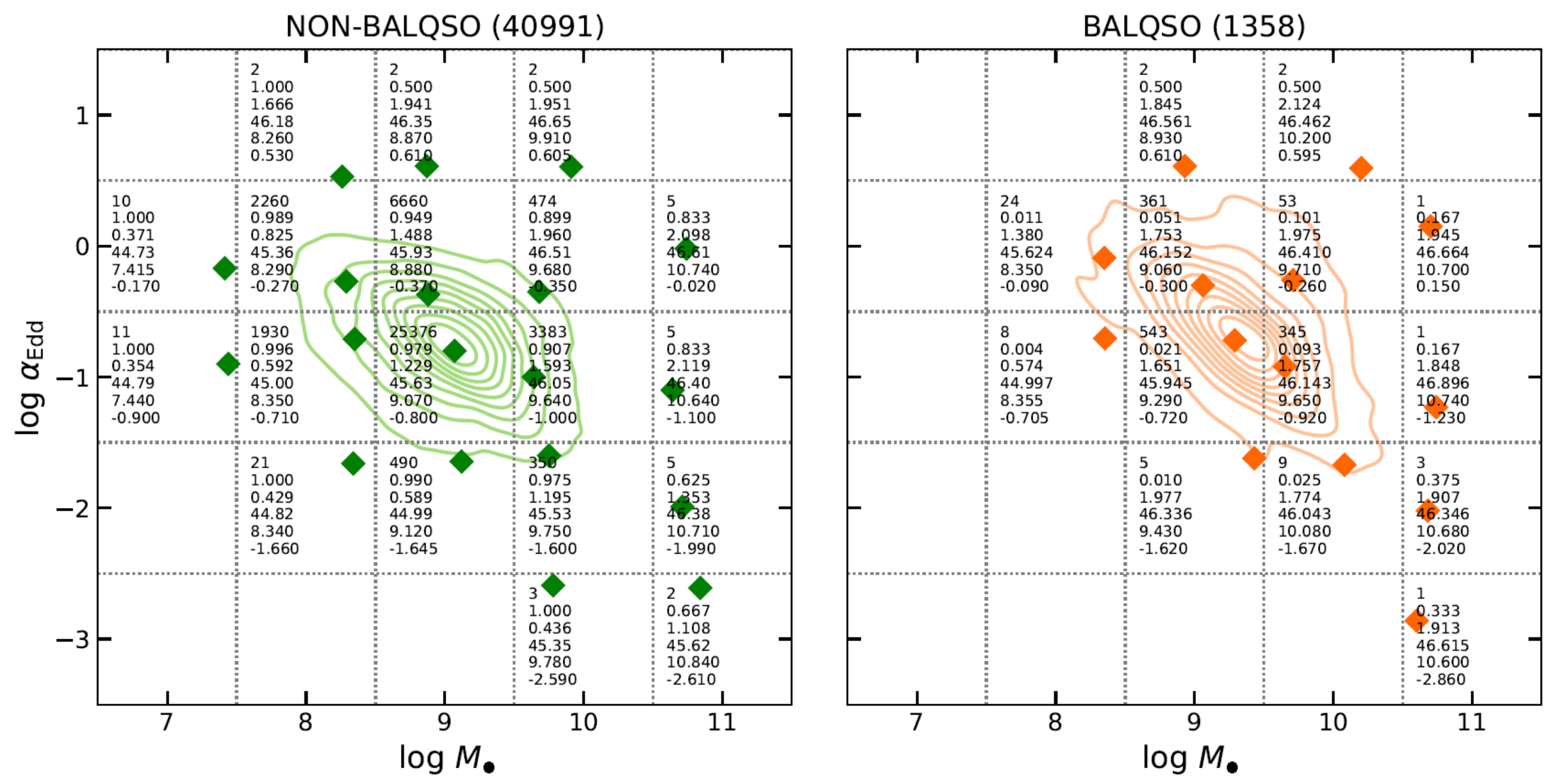}
        \caption{Black hole mass and  Eddington ratio space for non-BAL QSO (left panel) and BAL QSO (right panel). The space is divided into bins centred at $\Delta$log{$M_{\bullet}$}=[7.0, 8.0, 9.0, 10.0] and $\Delta$log\eddr=[-2.0, -1.0, 0.0]. The contours correspond to the full sample for each case, and the diamond symbols correspond to the median values of $\log{M_{\bullet}}$ and log \eddr\ in each bin. The number of sources, the ratio of non-BAL QSO (or BAL QSO) to the total sample size, and the median values of redshift, log$L_{3000}$, $\log{M_{\bullet}}$, and log\eddr\ are listed in each bin. }
        \label{fig:mbh-eddr_bins_mg2}
\end{figure*}

We show more examples (in fewer details, i.e. without a failed wind) in Figure~\ref{fig:one_Edd_outflow} for the Eddington rate. The trend with the black hole mass and metallicity is clear. Both affect the outflow very strongly. A higher metallicity means more dust content and a higher radiation pressure, so that the trend is rather natural. A higher Eddington rate also works in the same direction and enhances the radiation force. The trend with the black hole mass is also clear.
The change in outflow with the source parameters is extremely important in our model: for a low (solar) metallicity, an escaping zone forms only for higher black hole masses, and at high metallicity, the escape cone is much broader. 

All results are stored in Table~\ref{tab:compdata}. The absence of data in this table for some black hole masses, metallicities or Eddington rates means no escaping zone is formed for those parameters.

\begin{figure*}
  \centering  
  \includegraphics[width=0.8\textwidth]{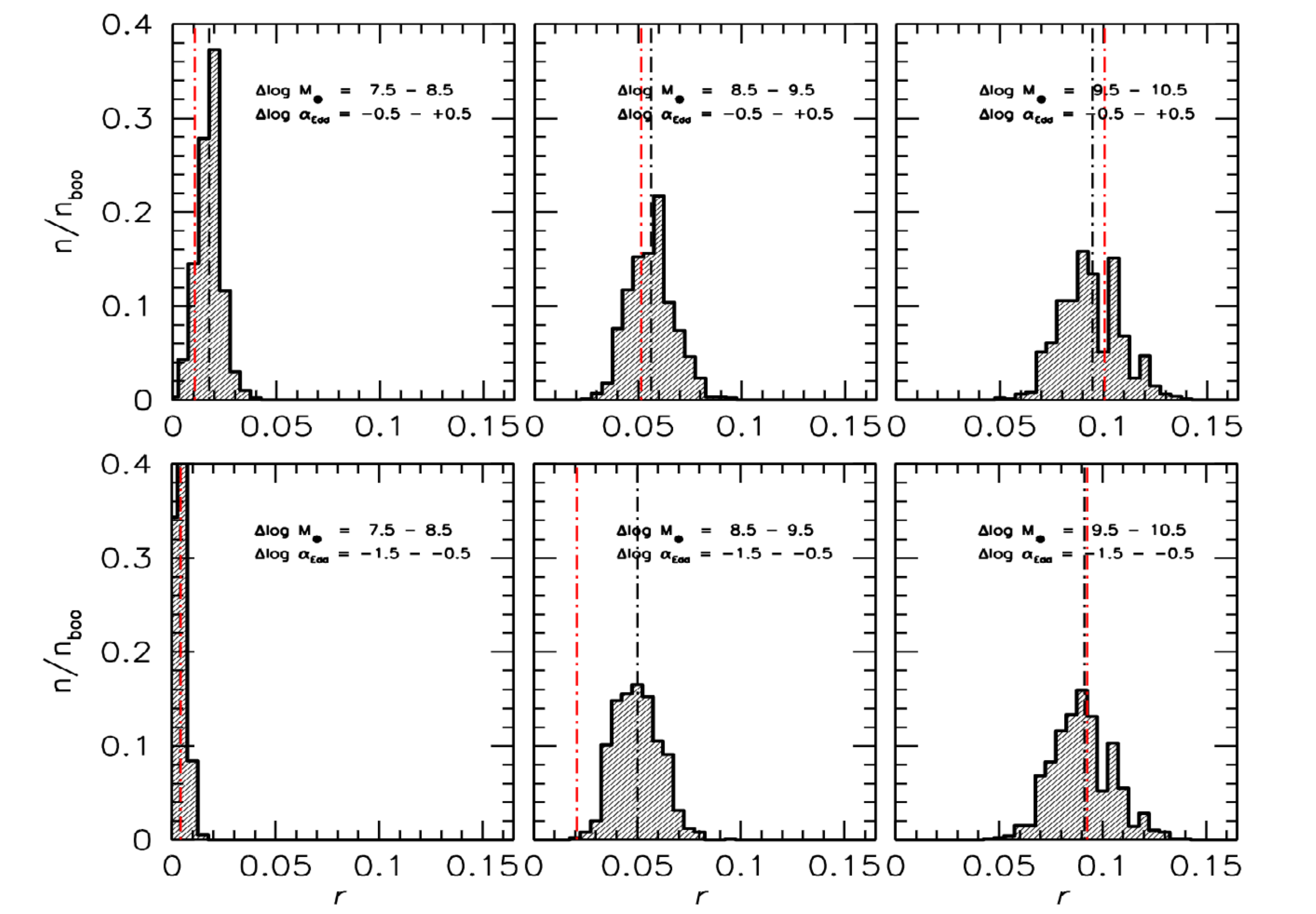}
    \caption{Distributions of the ratio $r = n_\mathrm{BAL}/n_\mathrm{Total}$ for 1000 bootstrap replications in six ranges limited in $M_{\bullet}$ and \eddr. The dot-dashed black line is the average of the bootstrapped distributions, and the red line traces the original $r$\ value from each bin sample.}
    \label{fig:boo}
\end{figure*}

\subsection{Analysis of the observational sample}

In order to compare the observational properties of BAL QSOs in our sample, we organised the results in two-dimensional space, parametrised by the black hole mass of the source and its Eddington rate for the \mgii\ emission line (see Fig.~{\ref{fig:mbh-eddr_bins_mg2}}). The total number of non-BAL QSOs with \mgiia mass measurements is large, 40 991 objects, and the number of BAL QSOs is 1358, high enough to study some trends. 
The black hole mass values, log $M_{\bullet}$, are predominantly located between 8 and 10, with the median values for non-BAL QSO close to $\log M_{\bullet} = 9.04$, and for BAL QSO close to $\log M_{\bullet} = 9.34$, a factor of 2 higher. A trend of the rising probability of the BAL phenomenon with the black hole mass is thus visible, which is statistically significant according to the KS test (see Sec.~\ref{sec:data}).} 

In order to compare the observational results with the theoretical models, the two-dimensional space occupied by the sample $M_{\bullet}$-\eddr\ was divided into bins centred at $\Delta \log{M_{\bullet}}$= [7.0, 8.0, 9.0, 10.0] and $\Delta$log{\eddr}=[-2.0, -1.0, 0.0]. Figure~\ref{fig:mbh-eddr_bins_mg2} shows the division of the $M_{\bullet}$-\eddr\ space, indicating the number of sources, redshift, luminosity at 3000$\rm \AA$, black hole mass, and Eddington ratio for non-BAL QSOs (left panel) and BAL QSOs (right panel) in each bin. Applying a KS test for redshift, luminosity, black hole mass, and Eddington ratio in each bin, we found $p$-values lower than $0.01$, which indicates a different parent distribution, as is the case of the full sample. For statistical consistency in each bin, we applied a bootstrap resampling (see Sec.~\ref{sec:bootstrap}). The number of BAL and non-BAL QSOs and the ratio of both  is listed in Table \ref{tab:nbal_fullsample}.

The contour lines show the object density distribution, and the overall BAL QSO distributions are not only shifted towards higher black hole masses, but are also more extended towards higher Eddington ratios, although the median for the Eddington rate is not much different. The distribution of BAL QSOs also seems to be more inclined with respect to the axis, that is, towards higher mass, lower Eddington ratio, and lower mass, higher Eddington ratio. When specific sub-types of BAL QSOs are analyzed, the distribution in two-dimensional space always seems to be more extended. 

\subsection{Boostrap resampling}
\label{sec:bootstrap}

\begin{figure*}
  \centering  
  \includegraphics[width=0.95\textwidth]{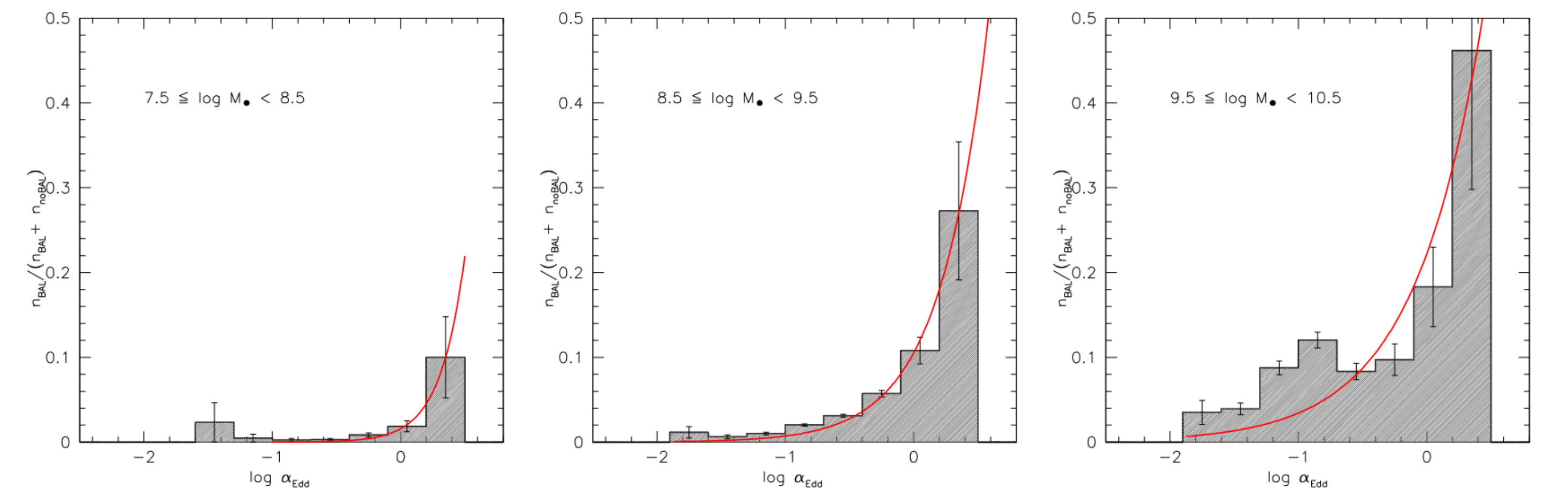}
    \caption{Distributions of the observational prevalence  ratio $r = n_\mathrm{BAL}/n_\mathrm{Total}$ as a function of the Eddington ratio for three $M_{\bullet}$ ranges. The error bars have been computed following Poissonian statistics.}
    \label{fig:redd}
\end{figure*}

We observe a trend in the sample that BAL QSOs seem to have a higher mass than non-BAL objects (see the $r$ value in Tab.~\ref{tab:nbal_fullsample}). In order to assess the statistical significance of this trend, we performed a bootstrap analysis.
Within each bin, a bootstrap resampling was carried out in bins in which the number of non-BAL and BAL QSO sources is higher than ten. One thousand simulated samples were computed, with the condition that the distributions of  $M_{\bullet}$ and \eddr\ did not differ significantly at the 2$\sigma$ \ confidence level and that any systematic difference between the median of $M_{\bullet}$ and \eddr\ distribution is lower than 0.1 dex. This was achieved by modelling the $M_{\bullet}$ and \eddr\ distributions of the non-BAL sources on the distribution of the BAL QSOs, and it accounted for any remaining bias in the distributions of the non-BAL QSOs and BAL QSOs. The effect is quite significant when the biases are strong, for example in the case of the 8.5 -- 9.5 $M_{\bullet}$ range, and --1.5 to --0.5 \eddr\ range. Figure \ref{fig:boo} shows a significant increase in the BAL prevalence as a function of $M_{\bullet}$: from the lowest bin centred at $\log{M_{\bullet}}$ $\sim$ 8 to the highest mass bin at $\log{M_{\bullet}}$ $\sim$ 10, there is a more than  fivefold increment for $-0.5 < \log \alpha_\mathrm{Edd} < 0.5$, and the increase is more than tenfold for $-1.55 < \log \alpha_\mathrm{Edd} < -0.5$. A separation of the BAL prevalence into two broad Eddington ratio ranges, --1.5 to --0.5, and --0.5 to +0.5, does not suggest any significant change in the prevalence as a function of \eddr. The BAL prevalence as a function of \eddr \ shows a strong trend. Figure \ref{fig:redd} shows a strong increase as a function of  \eddr\ for all the three mass ranges that reaches a maximum value $\approx 0.45$\ for the highest masses and for the highest \eddr\ in correspondence to the Eddington limit (right panel of Fig. \ref{fig:histograms}). The increase is well described by an exponential function, as detailed in Section \ref{sec:notorus}. In the distributions of Figure \ref{fig:boo}, the difference is washed away because only a few sources radiate close to the Eddington limit: for the range centred at $\log{M_{\bullet}}$ $\sim 10,$  only 5\%\ of all QSOs radiate this strongly (19 out of 407). 

\begin{table*}[h!]
    \caption{Observed BAL QSO  prevalences  as determined directly from the sample (original) and the corresponding distribution properties obtained from the bootstrap method (resampled)}
    \label{tab:nbal_fullsample}
    \begin{centering}
\begin{tabular}{ccccc|ccc}
\hline
\multicolumn{2}{c}{$\Delta\log{\alpha_{\mathrm{Edd}}}$} &
\multicolumn{3}{c}{-- 0.5 up to 0.5 } &
\multicolumn{3}{c}{-- 1.5 up to -- 0.5 } \\
\hline
\multicolumn{2}{c}{$\Delta\log{M_{\bullet}}$} & 7.5 -- 8.5  &   8.5 -- 9.5      &   9.5 -- 10.5   & 7.5 -- 8.5  &   8.5 -- 9.5    &   9.5 -- 10.5 \\
\hline\\
Original & $n_\mathrm{non-BAL}$ & $2260$  & $6660$  & $474$  & $1930$  & $25376$  & $3383$  \\
         & $n_\mathrm{BAL}$ & $24$  & $361$  & $53$  & $8$  & $543$  & $345$  \\
         & $r$ & $0.011$  & $0.051$  & $0.101$  & $0.004$  & $0.021$  & $0.093$  \\
         & $\mu_\frac{1}{2}$($\log{M_{\bullet}})_\mathrm{BAL}$ & $8.36$  & $9.06$  & $9.71$  & $8.38$  & $9.29$  & $9.65$ \\
         & $\delta \log{M_{\bullet}}$ & $0.07$ & $0.18$  & $0.03$  & $0.03$  & $0.22$  & $0.01$  \\
         & $\mu_\frac{1}{2}$($\log \alpha_\mathrm{Edd}$)$_\mathrm{BAL}$ & --$0.07$ & --$0.30$  & --$0.26$  & --$0.65$  & --$0.72$  & --$0.92$  \\
         & $\delta \log \alpha_\mathrm{Edd}$ & $0.20$ & $0.07$  & $0.09$  & $0.03$  & $0.08$  & $0.08$  \\
\\
\hline\\
Resampled & $r$ & $0.017$ & $0.056$  & $0.095$  & $0.004$  & $0.051$  & $0.091$  \\
          & $\sigma$ & $0.006$ & $0.011$  & $0.014$  & $0.0026$  & $0.010$  & $0.014$  \\
          &  $\mu_\frac{1}{2}$($\log{M_{\bullet}})_\mathrm{BAL}$ & $8.263$ & $9.085$  & $9.762$  & $8.250$  & $9.414$  & $9.701$  \\
          & $\delta \log{M_{\bullet}}$ & $0.069$ & $0.092$  & $0.029$  & --$0.052$  & $0.027$  & $0.022$  \\
          & $\mu_\frac{1}{2}$($\log \alpha_\mathrm{Edd}$)$_\mathrm{BAL}$ & $0.088$ & --$0.290$  & --$0.155$  & --$0.760$  & --$0.832$  & --$0.945$  \\
          & $\delta \log \alpha_\mathrm{Edd}$ & $0.103$ & $0.050$  & $0.043$  & --$0.001$  & $0.091$  & $0.052$  \\
\\
\hline
\end{tabular}
\\
\end{centering}
{$r=n_\mathrm{BAL}/n_\mathrm{Tot} = n_\mathrm{BAL}/(n_\mathrm{BAL}+n_\mathrm{non-BAL})$, $\mu_\frac{1}{2}$ corresponds to the median value in each case, $\delta \log{M_{\bullet}}$ and $\delta \log \alpha_\mathrm{Edd}$\ are defined as the median of BALs minus medians of non-BALs. Median values are shown in Fig. \ref{fig:boo}.}
\end{table*}

\begin{figure*}
  \centering  
  \includegraphics[width=\textwidth]{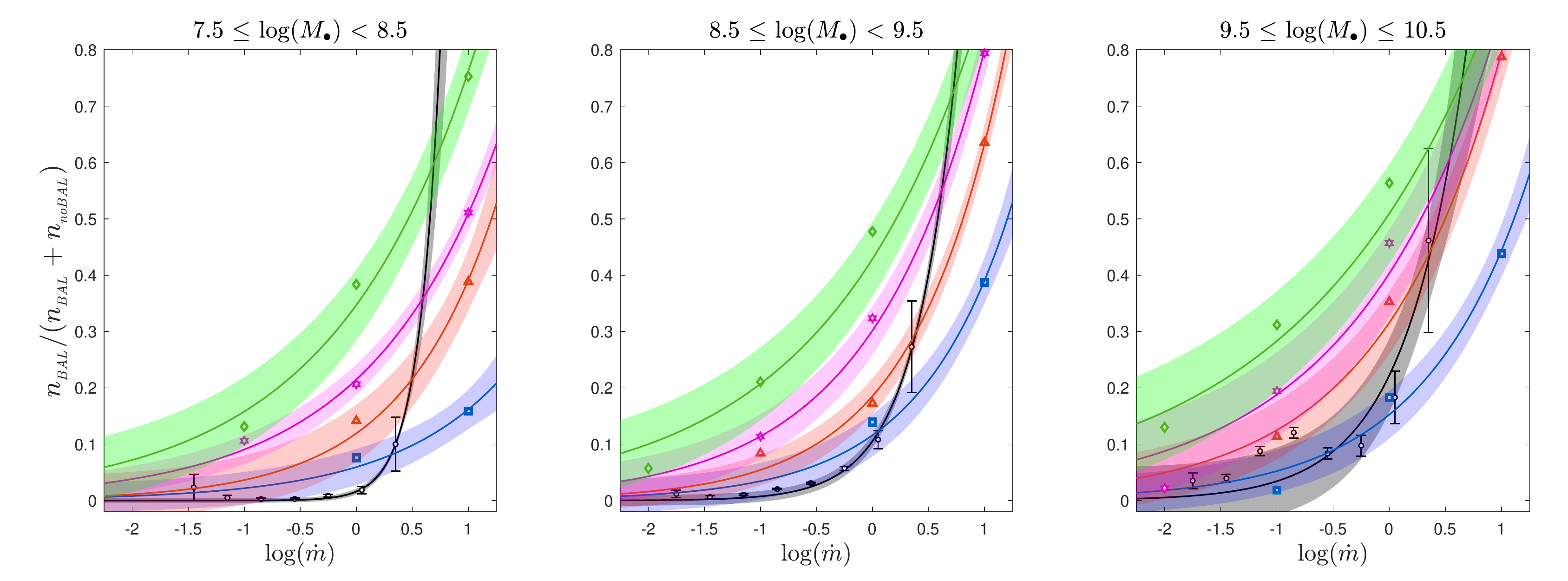}
    \caption{Distributions of the observational prevalence ratio $r = n_\mathrm{BAL}/n_\mathrm{Total}$ as a function of the Eddington ratio for three $M_{\bullet}$ ranges, shown by solid black lines, compared to simulation data for the case without a torus. The metallicities of 1, 2.5, 5, and 10 are depicted in blue, red, magenta, and green, respectively. The solid lines in each case show the best exponential fit to the corresponding data.}
    \label{fig:eddtrends}
\end{figure*}

\begin{figure*}
  \centering  
  \includegraphics[width=\textwidth]{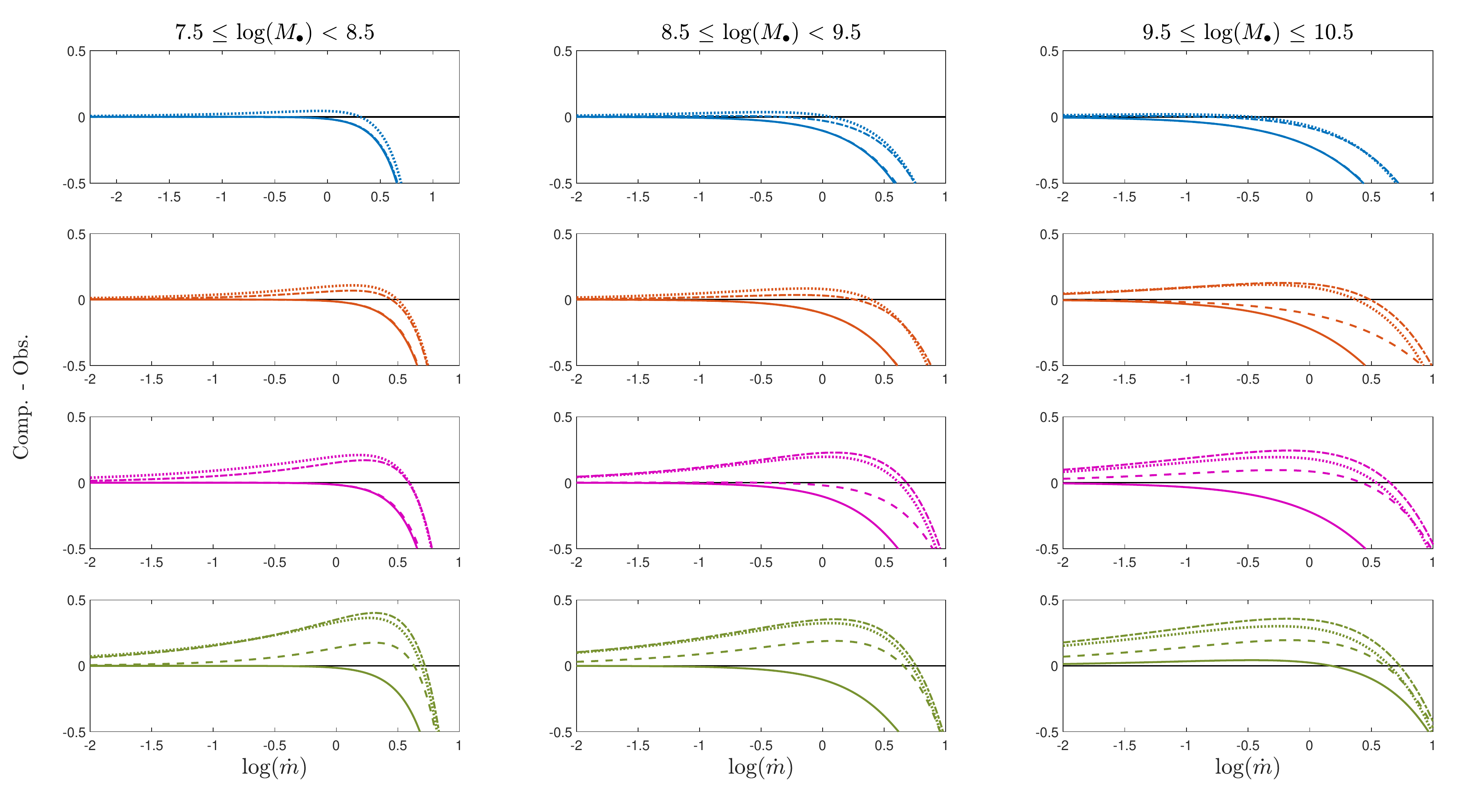}
    \caption{Residual of the computational results of the probability of observing BALs (simulations) subtracted from the prevalence ratio (observational data). The four adopted values of the metallicities of 1, 2.5, 5, and 10 are represented in blue, red, magenta, and green. The solid black line is depicted as a reference for the observational data. The solid, dashed, dash-dotted, and dotted lines correspond to opening angles of the torus of 45, 60, 75, and 90 (no torus), respectively. 
    The results are presented as a function of Eddington ratio (in steps of 0.3 dexes) for three $M_{\bullet}$ ranges.}
    \label{fig:balprevlog}
\end{figure*}

\subsection{Simulations compared with observational data}

\subsubsection{Without a torus}\label{sec:notorus}

The  observational data for the BAL prevalence ratio along with the simulation results of the BAL effect probability for cases without a torus ($\Omega_{\rm full}$, opening angle of 90  degrees) are shown in Figure~\ref{fig:eddtrends}, overplotted with the best exponential fit in each case. The results in Figure~\ref{fig:eddtrends} are colour-coded in black for the observational data and blue, red, magenta, and green for the simulation output with metallicities of 1, 2.5, 5, and 10, respectively. In all cases, the confidence intervals of 1$\sigma$ are shown as shaded belts embedding the best fit.

In general, the quantitative predictions of the FRADO model overlap with the data in the entire mass range. For low accretion rates, the solar metallicity is favoured, and for the Eddington rate, higher metallicities are better consistent with the model. This is an expected trend. The modelling of the broad emission lines in the optical plane also favours a higher metallicity for a higher Eddington rate \citep{martinez2018,panda2018, panda2019b, sniegowska2021, garnica2022}. No trend with the black hole mass is apparent: The required rise in the metallicity takes place in all cases above $\dot m \sim 3$, that is, only for super-Eddington sources, and this rise is not high. Metallicities of only up to 2 are required for $\dot m = 3$, although they rise steeply for still  higher Eddington rates.

\subsubsection{With a torus}\label{sec:withtorus}

If the torus is taken into account, the overall trend with the black hole mass remains the same, but the trend with metallicity is affected. To facilitate the comparison, the residual of the best-fit exponential functions for the computational results of the probability of observing BAL is subtracted by the exponential best fit for the prevalence ratio based on observational data and is presented in Figure~\ref{fig:balprevlog}. The same colour code as in Figure~\ref{fig:eddtrends} is followed for Figure~\ref{fig:balprevlog}, that is, the results for metallicities of 1, 2.5, 5, and 10 are represented in blue, red, magenta, and green, respectively.

For a thick torus (opening angle 45 deg), the results require a metallicity above 10 for an Eddington ratio above 1. For a thinner, less strongly shielding torus, the required metallicity is lower. Observations support a shielding torus. Its typical parameters  are not firmly established, but values of about 60 - 80 deg for the opening angle measured from the symmetry axis are usually derived \citep[][]{lawrence2010, alonso2011, Guainazzi2016, prince2022}. This would push our results based on BAL modelling toward metallicities of about a few times the solar value.

\section{Discussion} \label{sec:discussion}

We presented the first results of the comparison of a theoretical model, FRADO, with the probability of observing the BAL phenomenon based on the solid angle covered by the dust-driven outflow. The results are encouraging. The model predicts no BAL effect in sources with a low black hole mass. The outflow cone forms and increases with an increase in the black hole mass, Eddington rate, and metallicity. The trends predicted by the model are supported by a sample of BAL and non-BAL QSOs  from the SDSS DR7 Quasar catalogue \citep{shen2011}.

\subsection{Trends in the model and the data}

In the model, all three global parameters of mass, Eddington rate, and metallicity enhance the outflow and increase the fraction of the sky that is covered by the outflow from the point of view of the source. This increases the probability of observing the BAL phenomenon.

The selected subsample of non-BAL sources is large (40,991 objects). The non-BAL sources are less numerous (1,358 sources), but there are enough of them to see some trends. The increase in BAL probability with the black hole mass seems to be well visible. The trend with the Eddington ratio is also clear (see Figures~\ref{fig:redd} and \ref{fig:eddtrends}), and if all parameters are combined, we seem to see a need for higher metallicity in higher Eddington sources if the sources were to follow the model. Unfortunately, we do not have metallicity constraints for individual sources, only the mass and the Eddington ratio. However, this trend, a rise in metallicity with the Eddington rate, has frequently been concluded in independent studies \citep[e.g.][]{martinez2018,panda2018,panda2019b,sniegowska2021,shin2021, panda_pas,garnica2022}, although some studies instead reported a trend of the metallicity ratio with the black hole mass alone \citep[e.g.][]{matsuoka2011}.

We are unable to claim that the BAL phenomenon is not expected in low-mass AGN (log $M_{\bullet}\sim6-7$ $M_\odot$). The UV wavelength range is not covered in the SDSS DR7 observations. This does not permit us to identify BAL sources in low-mass AGNs. However, using observations from the Hubble Space Telescope and the International Ultraviolet Explorer, \citet{sulenticetal06a} reported the identification of 12 BAL QSOs at $0.05<z<0.46$ with an average mass of log $M_{\bullet}\sim8.1$ $M_\odot$, which is similar to the minimum value given by the SDSS DR7 quasar sample. In order to verify the presence of the low-mass BAL QSOs,  good UV data, particularly in a sample of high Eddington ratio AGNs, is needed.

\subsection{Possible role of the bias caused by the sample selection}

We selected suitable data to our best knowledge by exploring the effects in the sample associated with S/N  (Appendix~\ref{appendix: sn}) 
and with Malmquist-type biases (Appendix \ref{sect:appendix},  \ref{sect:appendixB}). Although there is a dependence on the redshift (luminosity) and the S/N that might affect the classification of a source (non-BAL or BAL QSO), we found no  significant difference  between the current classification of on-BAL and BAL QSOs according to a two-sample KS test (see Appendix~\ref{appendix: sn}).

On the other hand, we were unable to avoid some effects that are caused by the redshift because the identification of BALs depends on redshift.  Below $z = 1.5$, we cannot detect HBs because the CIV line is not covered. In order to gain insight into how this issue might affect the comparison between the data and the model, we studied our sample after the division into low- and high-redshift parts. The results are presented in Appendix~\ref{sect:appendix}. For the sources at a redshift above 1.5 alone, the probability of a BAL identification in sources with intermediate black hole masses is higher by a factor of 4. However, we lose any information about the low black hole mass sources (they are only detected at lower redshift in the current sample), which affects the dynamical range in the data and model comparison. The model predicts a strong trend with the black hole mass, and without a low-mass tail, we can only compare the results for black hole masses higher than $3 \times 10^8 M_{\odot}$. Redshift-limited samples still show the rise in the BAL fraction with mass, but the effect is not as strong as when the low-redshift and low-mass sources are included.

\subsection{Determination of the parameters in the quasar sample}

We selected only sources from \citet{shen2011} whose mass determination was based on \mgiia feature. Nevertheless, even in this case, the mass determination can have a considerable error, and this leads to an even large uncertainty in the derivation of the Eddington ratio because this requires the use of bolometric correction and an assumption that there is no intrinsic extinction of the continuum by dust. This might lead to large errors in some sources. However, statistically, the values of the mass and Eddington rate may not be strongly biased, although an improvement in the sample selection is possible in the future.

\subsection{Model assumptions}

The FRADO model allowed us to calculate the opening angle of the outflowing stream. It is based on a number of assumptions. The model assumes that the launch of the outflow is just due to radiation pressure that acts on dust. We discuss the potential problem with this assumption below. The model is not a hydro model, so that cloud collisions and cloud formation are not described, but the motion is highly supersonic, so that the effects of the pressure can be neglected. More advanced models can be built later, but this is not straightforward because of the extended character of the disk that powers the outflow and the need to include the dust evaporation role in the cloud dynamics.  

\subsubsection{Dust sublimation temperature}

The sublimation temperature of dust was set at 1500 K. Its precise value depends on the chemical composition of the dust as well as the grain size \citep[see][and the references therein]{baskin2018, panda2020a}. Selective dust evaporation is too complex to include in the present dynamical model. A global change in the sublimation temperature is only expected to radially shift the location of the escaping zone without affecting the orientation and opening angle of the outflow. 

\subsubsection{Outflow rate}

As the outflows can generally cause the disk to lose mass, which affects the accretion process, the assumption of a stationary disk with a radially constant accretion rate may not be correct.
On the other hand, much more material might flow out than finally accretes onto the central black hole because not only the kinetic luminosity was recently reported to be exceeding the bolometric luminosity of the quasar \citep{choi2020}, but estimates of the outflowing mass in a few sources also confirm this \citep{borguet2013, chamberlain2015}. Two scenarios can explain this issue: One scenario is based on assuming that the outflowing material is supplied from somewhere else than the disk itself, and in the second scenario, the outflow originates from the disk, but the accretion rate must be much higher at large radii \citep{naddaf2022MDPI}. Some estimates of the outflow within the frame of the 2.5D FRADO model were made in \citet{AnaLaura2022}, but they strongly depend on the additional assumptions because the model directly predicts the velocity, but not the density, of the wind or no of clouds per unit time. This effect is neglected in the current model.

\subsubsection{Line driving}

Our model is based on the assumption that the outflow is launched based on radiation pressure that acts on dust. Previous papers predicting the conical outflow were based on line-driving mechanisms. 

The formation of a fast funnel-shaped outflow of material from an accretion disk triggered by the disk radiation pressure was reported in many studies in hydro- and non-hydro contexts \citep[see e.g.][and the references therein]{proga2004, risaliti2010, Nomura2020}. The models are based on the line-driving mechanism, which is appropriate for the highly ionised part of the BLR located at a few hundred gravitational radii, $r_{\rm g}$ while the dust-driven outflow is launched at a few thousand  $r_{\rm g}$, which is more appropriate for a (partially) dusty outflow observed in BALs.

This does not mean that line driving is not important for BALs. The dust-driven outflow reaches only moderate velocities, while some measured BAL velocities approach a fraction of the speed of light, and may result from a line-locking mechanism \citep[e.g.][]{foltz1987,vilkoviskij2001,lin2020}. However, the outflow geometry is set in the first dust-driven stage, so that our approach should be a good approximation for the BAL geometry, the mean inclination of the stream and the covering factor, while detailed modelling of the outflow acceleration at later stages may require adding the line driving to the model. A combination of both line- and dust-driving mechanisms might help to explain the observed features of these objects \citep[see e.g.][]{marculewicz2022}.

\subsection{Implications for the torus}

Measurements of X-ray occultation indicate the presence of highly dynamical clouds, at all scales from the onset of BLR outwards to the torus, which obscure the X-ray source \citep{bianchi2012, markowitz2014, torricelli2014}.

Observational studies support the presence of the obscuring torus, with moderate up to very high opening angles \citep[][]{lawrence2010, alonso2011, Guainazzi2016, prince2022}. However, in excellent agreement with X-ray spectroscopy, a best-fit opening angle of $~ 50$ degrees is reported for the clumpy torus model \citep{nenkova2008a, nenkova2008b}.

Independent studies reported expectations for a higher metallicity in quasars when the Eddington ratio of the source increases \citep[e.g.][]{panda2018,panda2019b,sniegowska2021,shin2021,panda_pas, garnica2022}.

On the other hand, we simulated the motion of clumpy-dusty material lifted from the disk surface forming the BLR and outflowing towards the torus, and the results for the cases with the tours at moderate opening angles between 45 to 60 showed a better consistency with the observational data when higher metallicities were adopted. This can be considered as a relatively strong indication that the formation of a torus is a side effect of the formation of a BLR.

\section{Summary}

We tested the scenario according to which the BAL phenomenon in QSOs is not a temporary stage of their life. In this scenario, the BAL effect only occurs when the line of sight is within the spatially limited and collimated massive outflow (shallow) cone that covers only a fraction of the sky from the point of view of the nucleus. We used the theoretical model 2.5D FRADO, which predicts the cone geometry. The model is based on dust driving in the BLR region. 

We calculated the probability of observing a BAL QSO defined as the ratio of the solid angle of the outflow over the full sky, including considerations of the presence or absence of an obscuring torus. We compared our results with observational data for a sample of SDSS QSOs consisting of two sub-populations of BAL and non-BAL QSOs.  We found that in the model and in the data, the BAL phenomenon mostly occurs for sources whose black hole masses are higher than $10^8 M_{\odot}$, and the effect incrases with accretion rate. Moreover, high metallicities are more likely in QSOs showing BAL features when a torus is taken into account.

\begin{acknowledgements}
We thank the referee for the very helpful comments about the sample selection. The project was partially supported by the Polish Funding Agency National Science Centre, project 2017/26/A/ST9/00756 (MAESTRO 9), and MNiSW grant DIR/WK/2018/12. M. L. M.-A. acknowledges financial support from Millenium Nucleus NCN$19\_058$ (TITANs).  S.P. acknowledges the financial support from the Conselho Nacional de Desenvolvimento Científico e Tecnológico (CNPq) Fellowship (164753/2020-6). M.H.N. acknowledges the financial support for this research on the basis of a grant awarded in the 2022 competition for young scientists by the Center for Theoretical Physics, Polish Academy of Sciences. 
\end{acknowledgements}

\bibliographystyle{aa}
\bibliography{naddaf}

\begin{thebibliography}{80}
\expandafter\ifx\csname natexlab\endcsname\relax\def\natexlab#1{#1}\fi

\bibitem[{{Allen} {et~al.}(2011){Allen}, {Hewett}, {Maddox}, {Richards}, \&
  {Belokurov}}]{Allen_etal_2011}
{Allen}, J.~T., {Hewett}, P.~C., {Maddox}, N., {Richards}, G.~T., \&
  {Belokurov}, V. 2011, \mnras, 410, 860

\bibitem[{{Alonso-Herrero} {et~al.}(2011){Alonso-Herrero}, {Ramos Almeida},
  {Mason}, {Asensio Ramos}, {Roche}, {Levenson}, {Elitzur}, {Packham},
  {Rodr{\'\i}guez Espinosa}, {Young}, {D{\'\i}az-Santos}, \&
  {P{\'e}rez-Garc{\'\i}a}}]{alonso2011}
{Alonso-Herrero}, A., {Ramos Almeida}, C., {Mason}, R., {et~al.} 2011, \apj,
  736, 82

\bibitem[{{Baskin} \& {Laor}(2018)}]{baskin2018}
{Baskin}, A. \& {Laor}, A. 2018, \mnras, 474, 1970

\bibitem[{{Bautista} {et~al.}(2010){Bautista}, {Dunn}, {Arav}, {Korista},
  {Moe}, \& {Benn}}]{bautista2010}
{Bautista}, M.~A., {Dunn}, J.~P., {Arav}, N., {et~al.} 2010, \apj, 713, 25

\bibitem[{{Bianchi} {et~al.}(2012){Bianchi}, {Maiolino}, \&
  {Risaliti}}]{bianchi2012}
{Bianchi}, S., {Maiolino}, R., \& {Risaliti}, G. 2012, Advances in Astronomy,
  2012, 782030

\bibitem[{{Bischetti} {et~al.}(2022){Bischetti}, {Feruglio}, {D'Odorico},
  {Arav}, {Ba{\~n}ados}, {Becker}, {Bosman}, {Carniani}, {Cristiani}, {Cupani},
  {Davies}, {Eilers}, {Farina}, {Ferrara}, {Maiolino}, {Mazzucchelli},
  {Mesinger}, {Meyer}, {Onoue}, {Piconcelli}, {Ryan-Weber}, {Schindler},
  {Wang}, {Yang}, {Zhu}, \& {Fiore}}]{bischetti2022}
{Bischetti}, M., {Feruglio}, C., {D'Odorico}, V., {et~al.} 2022, \nat, 605, 244

\bibitem[{{Borguet} {et~al.}(2013){Borguet}, {Arav}, {Edmonds}, {Chamberlain},
  \& {Benn}}]{borguet2013}
{Borguet}, B. C.~J., {Arav}, N., {Edmonds}, D., {Chamberlain}, C., \& {Benn},
  C. 2013, \apj, 762, 49

\bibitem[{{Chamberlain} {et~al.}(2015){Chamberlain}, {Arav}, \&
  {Benn}}]{chamberlain2015}
{Chamberlain}, C., {Arav}, N., \& {Benn}, C. 2015, \mnras, 450, 1085

\bibitem[{{Choi} {et~al.}(2022{\natexlab{a}}){Choi}, {Leighly}, {Dabbieri},
  {Terndrup}, {Gallagher}, \& {Richards}}]{choietal22a}
{Choi}, H., {Leighly}, K.~M., {Dabbieri}, C., {et~al.} 2022{\natexlab{a}},
  \apj, 936, 110

\bibitem[{{Choi} {et~al.}(2022{\natexlab{b}}){Choi}, {Leighly}, {Terndrup},
  {Dabbieri}, {Gallagher}, \& {Richards}}]{choietal22}
{Choi}, H., {Leighly}, K.~M., {Terndrup}, D.~M., {et~al.} 2022{\natexlab{b}},
  \apj, 937, 74

\bibitem[{{Choi} {et~al.}(2020){Choi}, {Leighly}, {Terndrup}, {Gallagher}, \&
  {Richards}}]{choi2020}
{Choi}, H., {Leighly}, K.~M., {Terndrup}, D.~M., {Gallagher}, S.~C., \&
  {Richards}, G.~T. 2020, \apj, 891, 53

\bibitem[{{Czerny} {et~al.}(2016){Czerny}, {Du}, {Wang}, \&
  {Karas}}]{czerny2016}
{Czerny}, B., {Du}, P., {Wang}, J.-M., \& {Karas}, V. 2016, ApJ, 832, 15

\bibitem[{{Czerny} \& {Hryniewicz}(2011)}]{czerny2011}
{Czerny}, B. \& {Hryniewicz}, K. 2011, \aap, 525, L8

\bibitem[{{Czerny} {et~al.}(2017){Czerny}, {Li}, {Hryniewicz}, {Panda},
  {Wildy}, {Sniegowska}, {Wang}, {Sredzinska}, \& {Karas}}]{czerny2017}
{Czerny}, B., {Li}, Y.-R., {Hryniewicz}, K., {et~al.} 2017, ApJ, 846, 154

\bibitem[{{Czerny} {et~al.}(2015){Czerny}, {Modzelewska}, {Petrogalli}, {Pych},
  {Adhikari}, {{\.Z}ycki}, {Hryniewicz}, {Krupa}, {{\'S}wie{\c{t}}o{\'n}}, \&
  {Niko{\l}ajuk}}]{czerny2015}
{Czerny}, B., {Modzelewska}, J., {Petrogalli}, F., {et~al.} 2015, Advances in
  Space Research, 55, 1806

\bibitem[{{Di Matteo} {et~al.}(2005){Di Matteo}, {Springel}, \&
  {Hernquist}}]{DiMatteo2005}
{Di Matteo}, T., {Springel}, V., \& {Hernquist}, L. 2005, \nat, 433, 604

\bibitem[{{Drew} \& {Boksenberg}(1984)}]{drewboksenberg84}
{Drew}, J.~E. \& {Boksenberg}, A. 1984, \mnras, 211, 813

\bibitem[{{Dunn} {et~al.}(2010){Dunn}, {Bautista}, {Arav}, {Moe}, {Korista},
  {Costantini}, {Benn}, {Ellison}, \& {Edmonds}}]{dunn2010}
{Dunn}, J.~P., {Bautista}, M., {Arav}, N., {et~al.} 2010, \apj, 709, 611

\bibitem[{{Elvis}(2000)}]{elvis2000}
{Elvis}, M. 2000, \apj, 545, 63

\bibitem[{{Elvis}(2006)}]{Elvis2006}
{Elvis}, M. 2006, \memsai, 77, 573

\bibitem[{{Elvis}(2012)}]{Elvis2012radiation}
{Elvis}, M. 2012, in Astronomical Society of the Pacific Conference Series,
  Vol. 460, AGN Winds in Charleston, ed. G.~{Chartas}, F.~{Hamann}, \& K.~M.
  {Leighly}, 186

\bibitem[{{Foltz} {et~al.}(1987){Foltz}, {Weymann}, {Morris}, \&
  {Turnshek}}]{foltz1987}
{Foltz}, C.~B., {Weymann}, R.~J., {Morris}, S.~L., \& {Turnshek}, D.~A. 1987,
  \apj, 317, 450

\bibitem[{{Garnica} {et~al.}(2022){Garnica}, {Negrete}, {Marziani}, {Dultzin},
  {{\'S}niegowska}, \& {Panda}}]{garnica2022}
{Garnica}, K., {Negrete}, C.~A., {Marziani}, P., {et~al.} 2022, \aap, 667, A105

\bibitem[{{Gibson} {et~al.}(2009){Gibson}, {Jiang}, {Brandt}, {Hall}, {Shen},
  {Wu}, {Anderson}, {Schneider}, {Vanden Berk}, {Gallagher}, {Fan}, \&
  {York}}]{gibsonetal09}
{Gibson}, R.~R., {Jiang}, L., {Brandt}, W.~N., {et~al.} 2009, \apj, 692, 758

\bibitem[{{Gravity Collaboration} {et~al.}(2020){Gravity Collaboration},
  {Dexter}, {Shangguan}, {H{\"o}nig}, {Kishimoto}, {Lutz}, {Netzer}, {Davies},
  {Sturm}, {Pfuhl}, {Amorim}, {Baub{\"o}ck}, {Brandner}, {Cl{\'e}net}, {de
  Zeeuw}, {Eckart}, {Eisenhauer}, {F{\"o}rster Schreiber}, {Gao}, {Garcia},
  {Genzel}, {Gillessen}, {Gratadour}, {Jim{\'e}nez-Rosales}, {Lacour},
  {Millour}, {Ott}, {Paumard}, {Perraut}, {Perrin}, {Peterson}, {Petrucci},
  {Prieto}, {Rouan}, {Schartmann}, {Shimizu}, {Sternberg}, {Straub},
  {Straubmeier}, {Tacconi}, {Tristram}, {Vermot}, {Waisberg}, {Widmann}, \&
  {Woillez}}]{Gravity2020}
{Gravity Collaboration}, {Dexter}, J., {Shangguan}, J., {et~al.} 2020, \aap,
  635, A92

\bibitem[{{Guainazzi} {et~al.}(2016){Guainazzi}, {Risaliti}, {Awaki},
  {Arevalo}, {Bauer}, {Bianchi}, {Boggs}, {Brandt}, {Brightman}, {Christensen},
  {Craig}, {Forster}, {Hailey}, {Harrison}, {Koss}, {Longinotti}, {Markwardt},
  {Marinucci}, {Matt}, {Reynolds}, {Ricci}, {Stern}, {Svoboda}, {Walton}, \&
  {Zhang}}]{Guainazzi2016}
{Guainazzi}, M., {Risaliti}, G., {Awaki}, H., {et~al.} 2016, \mnras, 460, 1954

\bibitem[{{Hamann} {et~al.}(2019){Hamann}, {Herbst}, {Paris}, \&
  {Capellupo}}]{hamann2019}
{Hamann}, F., {Herbst}, H., {Paris}, I., \& {Capellupo}, D. 2019, \mnras, 483,
  1808

\bibitem[{{Hopkins} {et~al.}(2009){Hopkins}, {Murray}, \&
  {Thompson}}]{Hopkins2009}
{Hopkins}, P.~F., {Murray}, N., \& {Thompson}, T.~A. 2009, \mnras, 398, 303

\bibitem[{{Juarez} {et~al.}(2009){Juarez}, {Maiolino}, {Mujica}, {Pedani},
  {Marinoni}, {Nagao}, {Marconi}, \& {Oliva}}]{juarez2009}
{Juarez}, Y., {Maiolino}, R., {Mujica}, R., {et~al.} 2009, \aap, 494, L25

\bibitem[{{Knigge} {et~al.}(2008){Knigge}, {Scaringi}, {Goad}, \&
  {Cottis}}]{Knigge_etal_2008}
{Knigge}, C., {Scaringi}, S., {Goad}, M.~R., \& {Cottis}, C.~E. 2008, \mnras,
  386, 1426

\bibitem[{{Korista} {et~al.}(2008){Korista}, {Bautista}, {Arav}, {Moe},
  {Costantini}, \& {Benn}}]{korista2008}
{Korista}, K.~T., {Bautista}, M.~A., {Arav}, N., {et~al.} 2008, \apj, 688, 108

\bibitem[{{Krolik}(1999)}]{krolik_book_1999}
{Krolik}, J.~H. 1999, {Active Galactic Nuclei. From the Central Black Hole to
  the Galactic Environment}

\bibitem[{{Krolik} \& {Voit}(1998)}]{krolikvoit98}
{Krolik}, J.~H. \& {Voit}, G.~M. 1998, \apjl, 497, L5

\bibitem[{{Lawrence} \& {Elvis}(2010)}]{lawrence2010}
{Lawrence}, A. \& {Elvis}, M. 2010, \apj, 714, 561

\bibitem[{{Lin} \& {Lu}(2020)}]{lin2020}
{Lin}, Y.-R. \& {Lu}, W.-J. 2020, \mnras, 497, 1457

\bibitem[{{Lynds}(1967)}]{Lynds_1967}
{Lynds}, C.~R. 1967, \apj, 147, 396

\bibitem[{{Marculewicz} {et~al.}(2022){Marculewicz}, {Nikolajuk}, \&
  {R{\'o}{\.z}a{\'n}ska}}]{marculewicz2022}
{Marculewicz}, M., {Nikolajuk}, M., \& {R{\'o}{\.z}a{\'n}ska}, A. 2022, \aap,
  668, A128

\bibitem[{{Markowitz} {et~al.}(2014){Markowitz}, {Krumpe}, \&
  {Nikutta}}]{markowitz2014}
{Markowitz}, A.~G., {Krumpe}, M., \& {Nikutta}, R. 2014, \mnras, 439, 1403

\bibitem[{{Mart{\'\i}nez-Aldama} {et~al.}(2018){Mart{\'\i}nez-Aldama}, {del
  Olmo}, {Marziani}, {Sulentic}, {Negrete}, {Dultzin}, {D'Onofrio}, \&
  {Perea}}]{martinez2018}
{Mart{\'\i}nez-Aldama}, M.~L., {del Olmo}, A., {Marziani}, P., {et~al.} 2018,
  \aap, 618, A179

\bibitem[{{Mathis} {et~al.}(1977){Mathis}, {Rumpl}, \&
  {Nordsieck}}]{mathis1977}
{Mathis}, J.~S., {Rumpl}, W., \& {Nordsieck}, K.~H. 1977, \apj, 217, 425

\bibitem[{{Matsuoka} {et~al.}(2011){Matsuoka}, {Nagao}, {Marconi}, {Maiolino},
  \& {Taniguchi}}]{matsuoka2011}
{Matsuoka}, K., {Nagao}, T., {Marconi}, A., {Maiolino}, R., \& {Taniguchi}, Y.
  2011, \aap, 527, A100

\bibitem[{{Mej{\'\i}a-Restrepo} {et~al.}(2016){Mej{\'\i}a-Restrepo},
  {Trakhtenbrot}, {Lira}, {Netzer}, \& {Capellupo}}]{mejia-restrepo16}
{Mej{\'\i}a-Restrepo}, J.~E., {Trakhtenbrot}, B., {Lira}, P., {Netzer}, H., \&
  {Capellupo}, D.~M. 2016, \mnras, 460, 187

\bibitem[{{Moe} {et~al.}(2009){Moe}, {Arav}, {Bautista}, \&
  {Korista}}]{Moe2009}
{Moe}, M., {Arav}, N., {Bautista}, M.~A., \& {Korista}, K.~T. 2009, \apj, 706,
  525

\bibitem[{{M{\"u}ller} {et~al.}(2022){M{\"u}ller}, {Naddaf}, {Zaja{\v{c}}ek},
  {Czerny}, {Araudo}, \& {Karas}}]{AnaLaura2022}
{M{\"u}ller}, A.~L., {Naddaf}, M.-H., {Zaja{\v{c}}ek}, M., {et~al.} 2022, \apj,
  931, 39

\bibitem[{{Murray} {et~al.}(1995){Murray}, {Chiang}, {Grossman}, \&
  {Voit}}]{murray1995}
{Murray}, N., {Chiang}, J., {Grossman}, S.~A., \& {Voit}, G.~M. 1995, \apj,
  451, 498

\bibitem[{{Naddaf} \& {Czerny}(2022)}]{naddafczerny2022}
{Naddaf}, M.~H. \& {Czerny}, B. 2022, \aap, 663, A77

\bibitem[{{Naddaf} {et~al.}(2020){Naddaf}, {Czerny}, \&
  {Szczerba}}]{naddaf2020}
{Naddaf}, M.-H., {Czerny}, B., \& {Szczerba}, R. 2020, Frontiers in Astronomy
  and Space Sciences, 7, 15

\bibitem[{{Naddaf} {et~al.}(2021){Naddaf}, {Czerny}, \&
  {Szczerba}}]{naddaf2021}
{Naddaf}, M.-H., {Czerny}, B., \& {Szczerba}, R. 2021, \apj, 920, 30

\bibitem[{{Naddaf} {et~al.}(2022){Naddaf}, {Czerny}, \&
  {Zaja{\v{c}}ek}}]{naddaf2022MDPI}
{Naddaf}, M.-H., {Czerny}, B., \& {Zaja{\v{c}}ek}, M. 2022, Dynamics, 2, 295

\bibitem[{{Nenkova} {et~al.}(2008{\natexlab{a}}){Nenkova}, {Sirocky},
  {Ivezi{\'c}}, \& {Elitzur}}]{nenkova2008a}
{Nenkova}, M., {Sirocky}, M.~M., {Ivezi{\'c}}, {\v{Z}}., \& {Elitzur}, M.
  2008{\natexlab{a}}, \apj, 685, 147

\bibitem[{{Nenkova} {et~al.}(2008{\natexlab{b}}){Nenkova}, {Sirocky},
  {Nikutta}, {Ivezi{\'c}}, \& {Elitzur}}]{nenkova2008b}
{Nenkova}, M., {Sirocky}, M.~M., {Nikutta}, R., {Ivezi{\'c}}, {\v{Z}}., \&
  {Elitzur}, M. 2008{\natexlab{b}}, \apj, 685, 160

\bibitem[{{Nomura} {et~al.}(2020){Nomura}, {Ohsuga}, \& {Done}}]{Nomura2020}
{Nomura}, M., {Ohsuga}, K., \& {Done}, C. 2020, \mnras, 494, 3616

\bibitem[{{Nomura} {et~al.}(2013){Nomura}, {Ohsuga}, {Wada}, {Susa}, \&
  {Misawa}}]{nomura2013}
{Nomura}, M., {Ohsuga}, K., {Wada}, K., {Susa}, H., \& {Misawa}, T. 2013,
  \pasj, 65, 40

\bibitem[{{Panda} {et~al.}(2018){Panda}, {Czerny}, {Adhikari}, {Hryniewicz},
  {Wildy}, {Kuraszkiewicz}, \& {{\'S}niegowska}}]{panda2018}
{Panda}, S., {Czerny}, B., {Adhikari}, T.~P., {et~al.} 2018, \apj, 866, 115

\bibitem[{{Panda} {et~al.}(2020){Panda}, {Mart{\'\i}nez-Aldama}, {Marinello},
  {Czerny}, {Marziani}, \& {Dultzin}}]{panda2020a}
{Panda}, S., {Mart{\'\i}nez-Aldama}, M.~L., {Marinello}, M., {et~al.} 2020,
  \apj, 902, 76

\bibitem[{{Panda} {et~al.}(2019){Panda}, {Marziani}, \& {Czerny}}]{panda2019b}
{Panda}, S., {Marziani}, P., \& {Czerny}, B. 2019, \apj, 882, 79

\bibitem[{{Panda} \& {Skorek}(2022)}]{panda_pas}
{Panda}, S. \& {Skorek}, E.~J. 2022, in XL Polish Astronomical Society Meeting,
  ed. E.~{Szuszkiewicz}, A.~{Majczyna}, K.~{Ma{\l}ek}, M.~{Ratajczak},
  E.~{Niemczura}, U.~{B{\k{a}}k-St{\k{e}}{\'s}licka}, R.~{Poleski},
  M.~{Bilicki}, \& {\L}.~{Wyrzykowski}, Vol.~12, 72--75

\bibitem[{{Prince} {et~al.}(2022){Prince}, {Hryniewicz}, {Panda}, {Czerny}, \&
  {Pollo}}]{prince2022}
{Prince}, R., {Hryniewicz}, K., {Panda}, S., {Czerny}, B., \& {Pollo}, A. 2022,
  \apj, 925, 215

\bibitem[{{Proga} \& {Kallman}(2004)}]{proga2004}
{Proga}, D. \& {Kallman}, T.~R. 2004, \apj, 616, 688

\bibitem[{{Reichard} {et~al.}(2003){Reichard}, {Richards}, {Hall}, {Schneider},
  {Vanden Berk}, {Fan}, {York}, {Knapp}, \& {Brinkmann}}]{reichardetal03}
{Reichard}, T.~A., {Richards}, G.~T., {Hall}, P.~B., {et~al.} 2003, \aj, 126,
  2594

\bibitem[{{Risaliti} {et~al.}(2005){Risaliti}, {Bianchi}, {Matt}, {Baldi},
  {Elvis}, {Fabbiano}, \& {Zezas}}]{Risaliti2005}
{Risaliti}, G., {Bianchi}, S., {Matt}, G., {et~al.} 2005, \apjl, 630, L129

\bibitem[{{Risaliti} \& {Elvis}(2010)}]{risaliti2010}
{Risaliti}, G. \& {Elvis}, M. 2010, A\&A, 516, A89

\bibitem[{{Rodr{\'\i}guez Hidalgo} \& {Rankine}(2022)}]{Hidalgo2022}
{Rodr{\'\i}guez Hidalgo}, P. \& {Rankine}, A.~L. 2022, \apjl, 939, L24

\bibitem[{{R{\"o}llig} {et~al.}(2013){R{\"o}llig}, {Szczerba}, {Ossenkopf}, \&
  {Gl{\"u}ck}}]{szczerba2013}
{R{\"o}llig}, M., {Szczerba}, R., {Ossenkopf}, V., \& {Gl{\"u}ck}, C. 2013,
  \aap, 549, A85

\bibitem[{{Shen} {et~al.}(2011){Shen}, {Richards}, {Strauss}, {Hall},
  {Schneider}, {Snedden}, {Bizyaev}, {Brewington}, {Malanushenko},
  {Malanushenko}, {Oravetz}, {Pan}, \& {Simmons}}]{shen2011}
{Shen}, Y., {Richards}, G.~T., {Strauss}, M.~A., {et~al.} 2011, ApJs, 194, 45

\bibitem[{{Shen} {et~al.}(2019){Shen}, {Wu}, {Jiang}, {Ba{\~n}ados}, {Fan},
  {Ho}, {Riechers}, {Strauss}, {Venemans}, {Vestergaard}, {Walter}, {Wang},
  {Willott}, {Wu}, \& {Yang}}]{Shen_etal_2019}
{Shen}, Y., {Wu}, J., {Jiang}, L., {et~al.} 2019, \apj, 873, 35

\bibitem[{{Shin} {et~al.}(2021){Shin}, {Woo}, {Nagao}, {Kim}, \&
  {Bahk}}]{shin2021}
{Shin}, J., {Woo}, J.-H., {Nagao}, T., {Kim}, M., \& {Bahk}, H. 2021, \apj,
  917, 107

\bibitem[{{{\'S}niegowska} {et~al.}(2021){{\'S}niegowska}, {Marziani},
  {Czerny}, {Panda}, {Mart{\'\i}nez-Aldama}, {del Olmo}, \&
  {D'Onofrio}}]{sniegowska2021}
{{\'S}niegowska}, M., {Marziani}, P., {Czerny}, B., {et~al.} 2021, \apj, 910,
  115

\bibitem[{{Sulentic} {et~al.}(2006){Sulentic}, {Dultzin-Hacyan}, {Marziani},
  {Bongardo}, {Braito}, {Calvani}, \& {Zamanov}}]{sulenticetal06a}
{Sulentic}, J.~W., {Dultzin-Hacyan}, D., {Marziani}, P., {et~al.} 2006, Revista
  Mexicana de Astronomia y Astrofisica, 42, 23

\bibitem[{{Surdej} \& {Swings}(1981)}]{surdejswing81}
{Surdej}, J. \& {Swings}, J.~P. 1981, \aap, 96, 242

\bibitem[{{Tolea} {et~al.}(2002){Tolea}, {Krolik}, \&
  {Tsvetanov}}]{Tolea_etal_2002}
{Tolea}, A., {Krolik}, J.~H., \& {Tsvetanov}, Z. 2002, \apjl, 578, L31

\bibitem[{{Torricelli-Ciamponi} {et~al.}(2014){Torricelli-Ciamponi},
  {Pietrini}, {Risaliti}, \& {Salvati}}]{torricelli2014}
{Torricelli-Ciamponi}, G., {Pietrini}, P., {Risaliti}, G., \& {Salvati}, M.
  2014, \mnras, 442, 2116

\bibitem[{{Turnshek}(1988)}]{turnshek88}
{Turnshek}, D.~A. 1988, in QSO Absorption Lines: Probing the Universe, ed.
  J.~C. {Blades}, D.~A. {Turnshek}, \& C.~A. {Norman}, 17

\bibitem[{{Vestergaard}(2003)}]{vestergaard03}
{Vestergaard}, M. 2003, \apj, 599, 116

\bibitem[{{Vestergaard} \& {Peterson}(2006)}]{vestergaard2006}
{Vestergaard}, M. \& {Peterson}, B.~M. 2006, \apj, 641, 689

\bibitem[{{Vilkoviskij} \& {Irwin}(2001)}]{vilkoviskij2001}
{Vilkoviskij}, E.~Y. \& {Irwin}, M.~J. 2001, \mnras, 321, 4

\bibitem[{{Voit} {et~al.}(1993){Voit}, {Weymann}, \& {Korista}}]{voit1993}
{Voit}, G.~M., {Weymann}, R.~J., \& {Korista}, K.~T. 1993, \apj, 413, 95

\bibitem[{{Waters} {et~al.}(2021){Waters}, {Proga}, \& {Dannen}}]{waters2021}
{Waters}, T., {Proga}, D., \& {Dannen}, R. 2021, arXiv e-prints,
  arXiv:2101.09273

\bibitem[{{Weymann} {et~al.}(1991){Weymann}, {Morris}, {Foltz}, \&
  {Hewett}}]{weymannetal91}
{Weymann}, R.~J., {Morris}, S.~L., {Foltz}, C.~B., \& {Hewett}, P.~C. 1991,
  \apj, 373, 23

\bibitem[{{Young} {et~al.}(1999){Young}, {Corbett}, {Giannuzzo}, {Hough},
  {Robinson}, {Bailey}, \& {Axon}}]{young1999}
{Young}, S., {Corbett}, E.~A., {Giannuzzo}, M.~E., {et~al.} 1999, \mnras, 303,
  227

\end{thebibliography}


\begin{appendix}

\section{Dependence on  S/N}
\label{appendix: sn}

\begin{figure*}[ht!]
  \centering  
    \includegraphics[width=\textwidth]{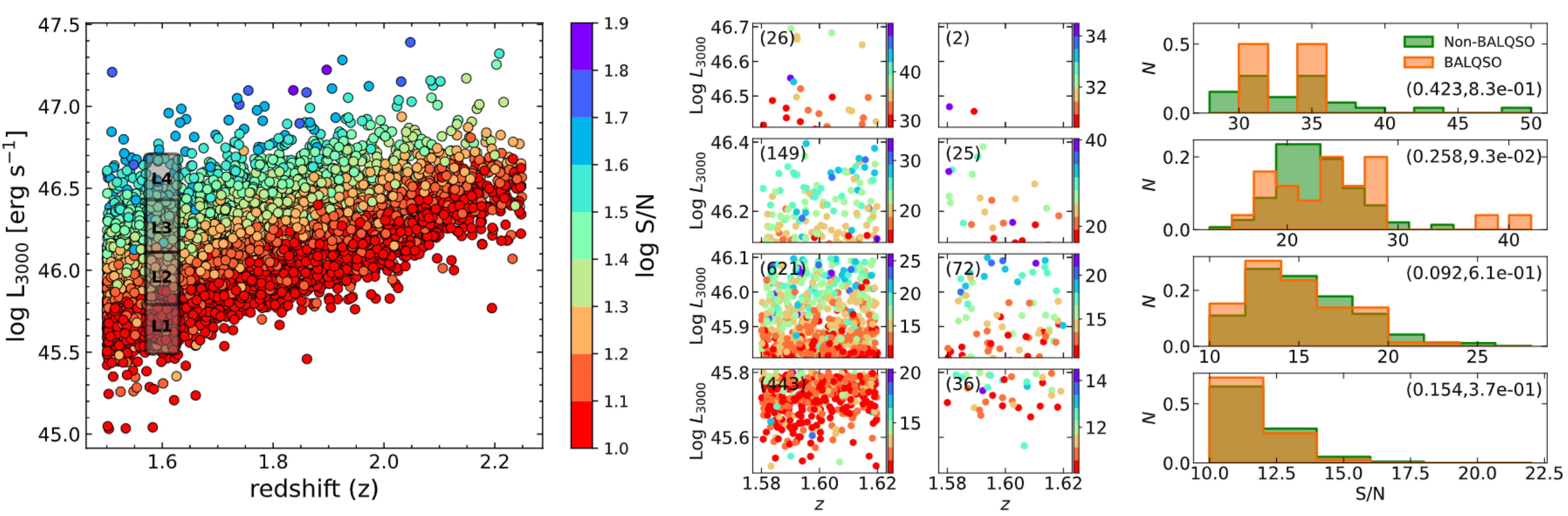}
    \caption{Bias analysis of the sample. Left panel: Relation between the luminosity at 3000\AA\ and redshift. The grey patch indicates the redshift bin we used to test the effect of the S/N (see Appendix~\ref{appendix: sn}). The colour code marks the S/N value in the log scale. Middle panel: Behavior of the S/N in each luminosity bin. The colour scale describes the S/N, and the number in parentheses corresponds to the number of sources in each bin. Right panel: Distribution of the S/N in each luminosity bin. The numbers in parentheses indicate the results from a two-sample KS test.}
    \vspace{20mm}
    \label{fig:z_L_sn}
\end{figure*}

One of the main biases in a sample is the spurious relation between the luminosity and the redshift (Malmquist bias), where the most luminous sources are most frequently found at high-$z$, and in turn, they show the highest S/N spectra. This is the case of the sample used in this analysis (see Fig.~\ref{fig:z_L_sn}, left panel), where a low S/N might underestimate the identification of low-luminosity BALAGN at low $z$. In order to test this effect, we selected a redshift bin centred at $z=1.6$ with a width of $\Delta z=0.05$ in the luminosity range of $45.5<L_{3000}<46.7$ with steps of $\Delta L_{3000}=0.3$ (see Fig.~\ref{fig:z_L_sn}). This binning covers the low-$z$ part of our sample showing luminosity and  S/N trends, such as the full redshift sample does (Fig.~\ref{fig:z_L_sn}, middle panel). Dividing the bin sample into BALAGNs and non-BALAGNs, we obtained the S/N distributions and performed a two-sample KS test in each luminosity bin (see Fig.~\ref{fig:z_L_sn}, right panel). According to the KS test, the non-BAL QSO and BAL QSO samples are drawn from the same distribution in all luminosity bins, being stronger in the lowest luminosity bins (L1 and L2) than in the highest luminosity bin (L4). However, the number of sources in the BAL QSO sample is too low in the highest luminosity bin (L4) to consider a proper statistic determination. These results indicate that the S/N does not affect the fraction of BAL QSO with respect to the full sample.

\section{Redshift-dependence effects in the prevalence of BAL QSOs}
\label{sect:appendix}

\begin{figure*}
  \centering  
  \includegraphics[width=0.80\textwidth]{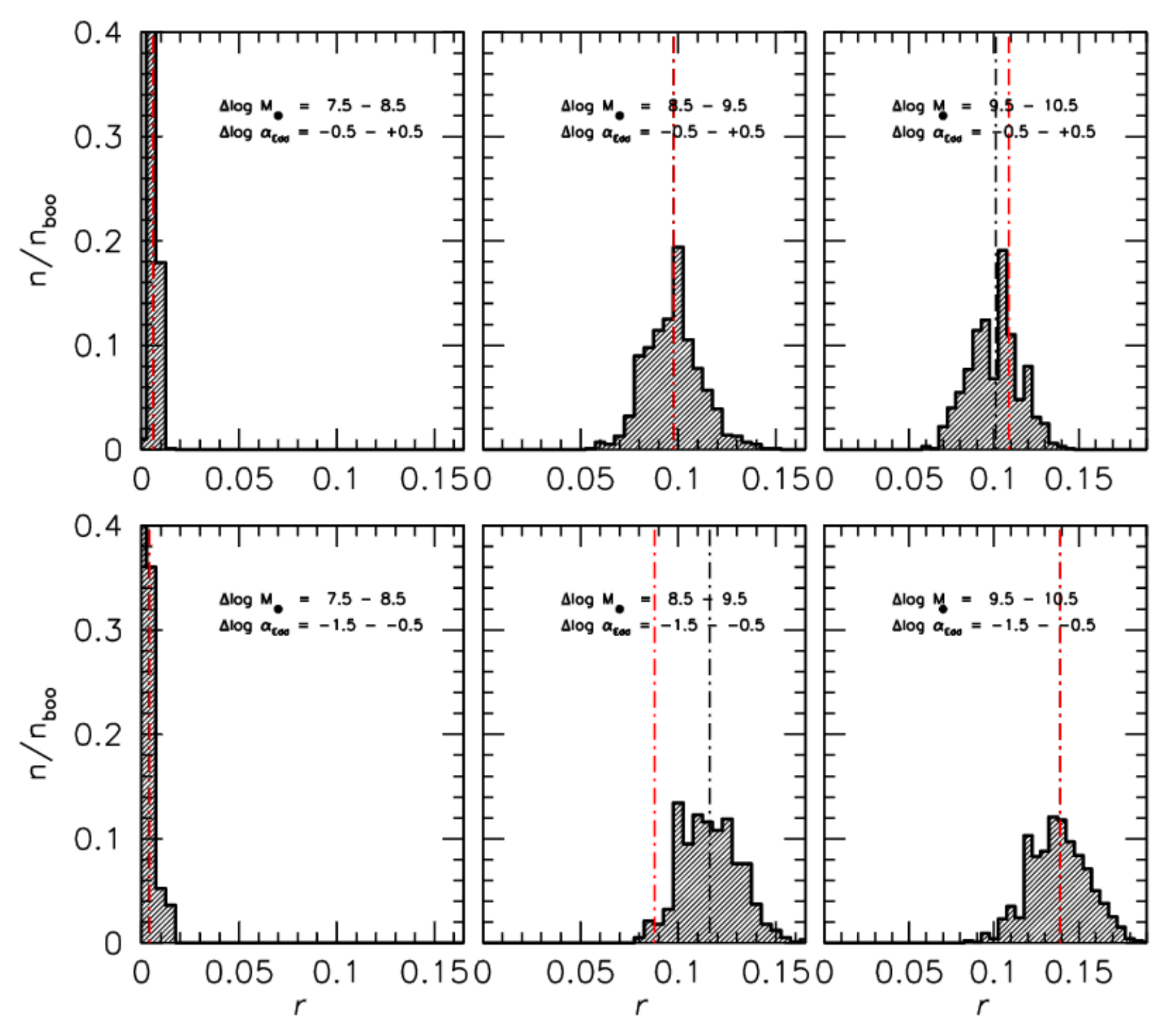}
    \caption{Distributions of the ratio $r = n_\mathrm{BAL}/n_\mathrm{Total}$ for 1000 bootstrap replications, in six ranges limited in $M_{\bullet}$ and \eddr, as in Fig. \ref{fig:boo}, but with restriction to $z \le 1.5$ (left panels) and $z > 1.5$\ (middle and right panels). The dot-dashed black line is the average of the bootstrapped distributions, and the red line traces the original $r$\ value from each bin sample.}
    \vspace{15mm}
    \label{fig:boonew}
\end{figure*}

The prevalence of BAL QSOs was recomputed by introducing a division of our sample into low- and high-redshift QSOs, assuming a redshift limit at $z = 1.5$, as shown in Fig. \ref{fig:boonew}. Below this limit, only LBs are sampled, while above this limit, both LBs and HBs are included, with LBs being a minority in the BAL population. This implies that the prevalences of BALs at $z \le 1.5$\ are by far lower ($\lesssim 0.01$) because almost only LBs are sampled. The prevalence of BALs is higher in the $\log$\mbh\ range centred at 9 and 10 because it is now computed excluding the large number of non-BALs at $z \le 1.5$\ where the majority HBs were entirely lost because \civ\ is not covered in the optical spectra. BAL QSOs are more strongly reddened in the rest-frame UV than non-BAL QSOs \citep{reichardetal03,gibsonetal09}. The difference in limiting magnitude due to the excess reddening was estimated as $\delta m \approx 0.15 - 0.2$, yielding an increase in the BAL fraction by $\approx 25$\%. This brings the estimates reported in Table \ref{tab:nbal_limitsample} ($\approx 10 - 14 $\%)  in reasonable agreement with previous estimates, 13 \% -- 17 \%\ (e.g. \citealt{gibsonetal09}).

\section{Trends as a function of Eddington ratio with redshift limits}
\label{sect:appendixB}

\begin{figure*}[hb!]
  \centering  
\includegraphics[width=1\textwidth]{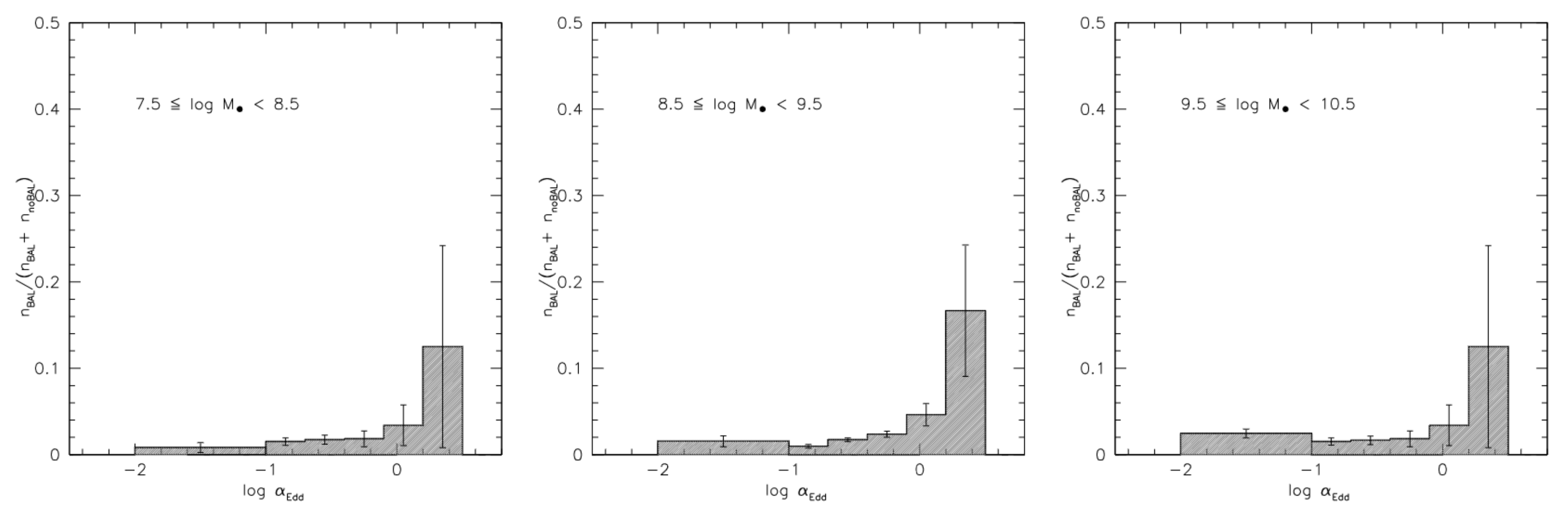}
    \caption{Distributions of the prevalence  ratio $r = n_\mathrm{BAL}/n_\mathrm{Total}$\ for only LBs as a function of Eddington ratio for three $M_{\bullet}$ ranges. The left panel is restricted to sources with $z \le 1.5$, and the middle and right panels show sources with $z > 1.5$.  The error bars have been computed following Poissonian statistics.}
    \label{fig:reddbin}
\end{figure*}

The FRADO predictions apply to LBs. We therefore restricted the relation between prevalence and $\alpha_\mathrm{Edd}$\ to only LBs. A trend involving a steep rise at the extreme $\alpha_\mathrm{Edd}$\ is still visible, although   small number statistics make the results uncertain, except in the range centred at $\log$\mbh = 9\ (Fig.  \ref{fig:reddbin}, middle panel). An increase in the prevalence at the low end of $\alpha_\mathrm{Edd}$\ may be due to small number statistics, as it involves just one or two objects. We therefore extended the first bin range to $-2 \lesssim \alpha_\mathrm{Edd} <  -1$.

\begin{table*}[hb!]
    \caption{Observed BAL QSO  prevalences  as determined from the samples with a limit at $z = 1.5$, and the corresponding distribution properties obtained from the bootstrap method (resampled)}
    \label{tab:nbal_limitsample}
    \begin{centering}
\begin{tabular}{ccccc|ccc}
\hline
\multicolumn{2}{c}{$\Delta\log{\alpha_{\mathrm{Edd}}}$} &
\multicolumn{3}{c}{-- 0.5 ~up to~ 0.5 } &
\multicolumn{3}{c}{-- 1.5 ~up to~ -- 0.5 } \\
\hline
\multicolumn{2}{c}{$\Delta\log{M_{\bullet}}$} & 7.5 -- 8.5  &   8.5 -- 9.5      &   9.5 -- 10.5   & 7.5 -- 8.5  &   8.5 -- 9.5    &   9.5 -- 10.5 \\
\hline
\multicolumn{2}{c}{$\Delta z$} &
0 -- 1.5 & $> 1.5$ & $> 1.5$ & 0 -- 1.5 & $> 1.5$ &$> 1.5$ \\ 
\hline \\
          Original & $n_\mathrm{non-BAL}$                                         & $2159$  &     $3211 $ & $434$               &  $1927$            & $5279$      & $2098$        \\
         &  $n_\mathrm{BAL}$                                                                                                   & $13$     &    $348$  &  $53$               &  $8$               & $508$       & $339$         \\
         & $r$                                                                  & $0.006$  &    $0.098$    & $0.109$          &  $0.004$           & $0.088$     & $0.139$        \\
         & $\mu_\frac{1}{2}$($\log{M_{\bullet}})_\mathrm{BAL}$                      & $8.20$   &    $9.08$  &  $9.71$              &  $8.38$             & $9.30$      & $9.65$        \\
         & $\delta \log{M_{\bullet}}$                                           & $-0.08$  &   $0.07$  & $0.03$          &    $   0.03$             & $0.06$     & $-0.01$        \\
         & $\mu_\frac{1}{2}$($\log \alpha_\mathrm{Edd}$)$_\mathrm{BAL}$         & --$0.17$ &  --$0.29$  & --$0.65$             &  --$0.65$          & --$0.72$    & --$0.92$        \\
         & $\delta \log \alpha_\mathrm{Edd}$                                    & $0.11$  &  $0.05$   & $0.06$              &  $0.06$           & $-0.02$     & $0.00$        \\ \\ \hline
         \\
        Resampled       & $r$                                                      & $0.0063$  &    $0.098$    & $0.101$           &  $0.004$           & $0.116$     & $0.139$        \\
         & $\mu_\frac{1}{2}$($\log{M_{\bullet}})_\mathrm{BAL}$                      & $8.18$   &     $9.066$     &  $9.75$              &  $8.215$             & $9.4$      & $9.70$        \\
         & $\delta \log{M_{\bullet}}$                                           & $-0.028$  &    $0.04$    & $0.003$             & $  -0.083$             & $0.008$     & $0.004$        \\
         & $\mu_\frac{1}{2}$($\log \alpha_\mathrm{Edd}$)$_\mathrm{BAL}$         & --$0.08$ &     $-0.276$    & --$0.157$             &  --$0.68$          & --$0.798$    & --$0.93$        \\
         & $\delta \log \alpha_\mathrm{Edd}$                                    & $0.076$  &    $0.037$    & $-0.004$              &  $0.068$           & $0.005$     & $-0.003$        \\
                \\
\hline
\end{tabular}
\\
\end{centering}
{$r=n_\mathrm{BAL}/n_\mathrm{Tot} = n_\mathrm{BAL}/(n_\mathrm{BAL}+n_\mathrm{non-BAL})$, $\mu_\frac{1}{2}$ corresponds to the median value in each case, $\delta \log{M_{\bullet}}$ and $\delta \log \alpha_\mathrm{Edd}$\ are defined as the median of BALs minus medians of non-BALs. Median values are shown in Fig. \ref{fig:boonew}.}
\end{table*}

\end{appendix}

\end{document}